\documentclass[prc,aps,floatfix,nofootinbib,twocolumn,letterpaper]{revtex4}
\usepackage{epsfig} 
\usepackage{amssymb} 
\usepackage{amsmath} 
\usepackage{amsfonts} 
\usepackage{color,graphicx} 
\usepackage{bm}
\def\Mevinv{MeV$^{-1}$}

\begin{document} 
\title{Benchmarking mean-field approximations to level densities} 
\author{Y. Alhassid,$^{1}$, G.F.~Bertsch$^{2}$, C.N.~Gilbreth$^{1}$, and 
H. Nakada$^3$  
} 
\affiliation{$^{1}$Center for Theoretical Physics, Sloane Physics 
Laboratory, Yale University, New Haven, Connecticut 06520, USA\\ 
$^{2}$Department of Physics and Institute for Nuclear Theory, 
Box 351560\\ University of Washington, Seattle, Washington 98915, USA\\
$^3$Department of Physics, Graduate School of Science,Chiba University, Inage, Chiba 263-8522, Japan} 
\def\lb{\langle} 
\def\rb{\rangle} 
\def\ni{\noindent} 
\def\be{\begin{equation}} 
\def\ee{\end{equation}} 
\def\sumk{\sum_k} 
\def\ad{a^\dagger_k} 
\def\adb{a^\dagger_{\bar k}} 
\def\a{a_k} 
\def\ab{a_{\bar k}} 
\def\Tr{{\rm Tr}} 
\def\tr{{\rm tr}} 
\def\Re{{\rm Re\,}} 
\def\sm{$^{148}$Sm} 
\def\dy{$^{162}$Dy} 
 
\begin{abstract} 

We assess the accuracy of finite-temperature mean-field theory
using as a standard the Hamiltonian and model space of the shell model Monte Carlo
calculations. 
Two examples are considered: the nucleus $^{162}$Dy, 
representing a heavy deformed nucleus, and  $^{148}$Sm, representing a nearby heavy
 spherical nucleus with strong pairing correlations.  The errors 
inherent in the finite-temperature Hartree-Fock and Hartree-Fock-Bogoliubov approximations are 
analyzed by comparing the entropies of the grand canonical and canonical 
ensembles, as well as the level density at the neutron resonance threshold, with shell model Monte Carlo (SMMC) calculations, which are accurate up to well-controlled statistical errors.  The main 
weak points in the mean-field treatments are seen to be: (i) the extraction 
of number-projected densities from the grand canonical ensembles, and (ii) the 
symmetry breaking by deformation or by the pairing condensate. In the absence of a pairing condensate, 
we confirm that the usual saddle-point approximation to extract the 
number-projected densities is not a significant source of error 
compared to other errors inherent to the mean-field theory.  We also present an 
alternative formulation of the saddle-point approximation that makes direct 
use of an approximate particle-number projection and avoids computing the usual 
three-dimensional Jacobian of the saddle-point integration.  We 
find that the pairing condensate is less amenable to approximate 
particle-number projection methods due to the explicit violation of particle-number 
conservation in the pairing condensate.  Nevertheless, the Hartree-Fock-Bogoliubov theory 
is accurate to less than one unit of entropy for $^{148}$Sm at 
the neutron threshold energy, which is above the pairing phase transition.  
This result provides support for the commonly used ``back-shift''
approximation, treating pairing as only affecting the excitation energy 
scale.  When the ground state is strongly deformed, the Hartree-Fock entropy 
is significantly lower than the SMMC entropy at low temperatures due to the missing contribution of rotational 
degrees of freedom.  However, treating the rotational bands in a simple model, we find that 
the entropy at moderate excitation energies is reproduced to within two units, 
corresponding to an error in the level density of less than an order of 
magnitude.  We conclude with a discussion of methods that have 
been advocated as beyond the mean field approximation, and their prospects 
to ameliorate the issues we have identified. 
\end{abstract} 
 
 
\maketitle 
 
\section{Introduction} 
 
Nuclear level densities are important input in the theory 
of low-energy nuclear reactions. In situations where the reactions cannot  
be studied in the laboratory, the need for a reliable theory 
is evident. The calculation of level densities in the presence of correlations is a challenging problem. Part of the problem is the complexity  
and inadequate knowledge of the nuclear Hamiltonian.  But even with 
a known Hamiltonian, it is often necessary to make approximations based on mean-field theory whose accuracy is not well understood.
 Our goal here  is to assess these mean-field approximations by taking the Hamiltonian 
as known and testing them against a theory that is accurate up to 
well-controlled statistical errors.  
 
Most treatments of the level density of heavy nuclei start from a mean-field
theory in the form of the finite-temperature Hartree-Fock (HF) or the
Hartree-Fock-Bogoliubov (HFB) approximation.  The finite-temperature theory
is derived from a variational principle based on 
the grand potential as a function of an uncorrelated trial density~\cite{go81,ta81}.  These approximations have been
widely applied and taken as a starting point for more sophisticated
theories~\cite{ke81,va83}.  It would thus be useful to validate them by
comparing with a more accurate method.  The auxiliary-field Monte Carlo
method, known as the shell model Monte Carlo (SMMC) in nuclear
physics~\cite{la93,al94}, fulfills this role.  Given the Hamiltonian and a model
space, the only inaccuracy is a controllable statistical error associated
with the Monte Carlo sampling.  As a finite-temperature method, SMMC 
is particularly suitable for the calculation of level densities~\cite{na97,al99}.
 
The SMMC starts from a Hamiltonian that is somewhat restricted but which
reproduces all the long-range correlations in a realistic way.  These include
deformations, pairing gaps and low-energy collective excitations.  It is
thus well-suited to provide a benchmark for testing the validity of the
mean-field treatments of nuclear thermal and statistical properties. We note
that the SMMC applies to all nuclei irrespective of whether they are
deformed or spherical and independently of the existence of a mean-field
pairing condensate. In contrast, the formulas for level densities in HF and
HFB depend on the character of the mean-field solution.

In Sec.~\ref{statistical} we summarize the statistical and thermodynamic
tools we will use for the analysis and comparisions.  In Sec.~\ref{SMMC} 
we present in some detail the results of the SSMC calculations for
two nuclei, namely \sm~and \dy.  The first is
a spherical nucleus having a strong pairing condensate in HFB.  The second
is a well-deformed nucleus with a weak pairing condensate; here the HF is
an appropriate mean-field approximation.
In Sec.~\ref{HF} we present the HF approximation and its results  for \dy,
while in Sec.~\ref{HFB} we discuss the HFB approximation and its results for \sm. 
In Sec.~\ref{conclusion} we summarize our
findings regarding the accuracy of the HF and HFB approximations with some
remarks on prospects for extending the mean-field approach to level densities.
Finally, the data files from the SMMC, HF and HFB calculations together with
the codes to apply them are provided in the Supplementary Material
depository accompanying this article. 

\section{Tools of statistical theory}\label{statistical}
\label{basics}
\subsection{The canonical entropy}
\label{state-density}
A good meeting point for comparing different statistical theories is the canonical
entropy function $S_c(\beta,N_p,N_n)$,  i.e., the entropy of the 
canonical ensemble of states having fixed numbers of protons and neutrons  $N_p, N_n$ and 
at inverse temperature $\beta$.  
In the SMMC, the quantity
that is most directly computed is the thermal energy $E_c(\beta)$ of 
the canonical ensemble.
The entropy can then computed by 
integrating the thermodynamic relation 
\be
\label{dS}
dS_c  = \beta\, dE_c
\ee
as
\be 
\label{S}
S_c(\beta) = S_c(0) -\int_{E_c(\beta)}^{E_c(0)} \beta'\, d E_c \;.
\ee  
This allows one to calculate the partition function $Z_c$ from
$Z_c = \exp(-\beta E_c + S_c)$.  Alternatively, $Z_c$ can be calculated directly from $E_c$
by integrating the relation 
\be
d \ln Z_c = -E_c d\beta \;,
\ee
and the canonical entropy is then calculated from $S_c = \ln Z_c +\beta E_c$. 
Finally, the density of states 
$\rho(E,N_p,N_n)$ at given energy $E$ and particle numbers $N_p, N_n$ is obtained from $Z_c$ by an inverse Laplace transform,
carried out in the saddle-point approximation
\begin{eqnarray}\label{s2r}
\rho(E,N_p,N_n) & = &
{1\over 2\pi i} \int^{i\infty}_{-i\infty}
d \beta'\, e^{\beta' E} Z_c \nonumber \\
&\approx & \left(2 \pi \left|{\partial  E \over\partial \beta}\right|
\right)^{-1/2} e^{S_c(\beta)} \;,
\end{eqnarray}
where $\beta$ in the above expression is determined as a function of $E$ from the saddle-point condition
\be
 E = - {\partial \ln Z_c \over \partial \beta}= E_c(\beta) \;.
\ee

The entropy of a system whose Hamiltonian is defined in a finite-dimensional model space satisfies a sum rule 
obtained from (\ref{S}) in the limit $\beta \to \infty$
\be
\label{sr}
\int_{E(\beta=\infty)}^{E(\beta=0)} \beta\, d E  = S(0) - S(\infty).
\ee  
For a finite-dimensional model space, the entropy at both end points $\beta=0$ and $\beta=\infty$ 
is finite and can be determined analytically.
In particular, for even-even nuclei, the entropy at zero temperature $S_c(\infty)$ must be zero. 

In the configuration-interaction (CI) shell model, we have a certain number $N_p,N_n$ of nucleons in single-particle model 
spaces of dimensions $D_p, D_n$ giving a canonical entropy at $\beta=0$ of
\be\label{Sc0} 
S_c(0) = S_p(0)+S_n(0) = \ln {D_p \choose N_p}+\ln {D_n \choose 
N_p}. 
\ee 

We have found  the sum rule in Eq.~(\ref{sr}) useful for testing
the computer codes we have employed in this study, and also
for setting end points on the entropy plots we show later.

\subsection{The grand canonical entropy}

The finite-temperature HF (FTHF) and finite-temperature HFB (FTHFB) approximations are defined in the framework of the grand canonical ensemble which depends on the additional independent variables $\alpha_i$ ($i=p,n$), related to the chemical potentials $\mu_i$ by $\mu_i = \alpha_i/\beta$.  

The grand canonical entropy $S_{gc}$, when expressed as a function of the energy $E_{gc}$ and the  average number of particles $N_{i,gc}$ of  type $i$ in the grand canonical ensemble, satisfies
\be
d S_{gc} = \beta d E_{gc} - \sum_{i=p,n} \alpha_i dN_{i,gc} \;,
\ee
and can be calculated as in Eq.~(\ref{S}) for fixed $N_{p,gc},N_{n,gc}$, i.e., when the grand canonical energy $E_{gc}$ is expressed as a function of $\beta$ and the given particle numbers $N_{i,gc}$.

The grand canonical  entropy at $\beta = 0$ 
is also a simple function of the dimension of the single-particle shell-model space in the CI shell model,
since all correlations disappear in that limit. Choosing values for $\alpha_p=\beta \mu_p,\alpha_n=\beta\mu_n$ (where $\mu_p,\mu_n$ are chemical potentials) 
to produce average particle numbers $N_p,N_n$, we have
\be 
\label{Sgci0} 
S_{gc}(0,N_p,N_n) = - \sum_{i=p,n}D_i \left[ f_i \ln f_i + (1-f_i) \ln 
(1-f_i)\right] \;, 
\ee 
where $f_i = N_i/D_i$ are occupation factors for $i=p,n$.  

 The grand canonical partition $Z_{gc}=Z_{gc}(\beta,\alpha_p,\alpha_n)$ satisfies
\be
d \ln Z_{gc}= -E_{gc} d\beta +\sum_i  N_{i,gc} d\alpha_i \;.
\ee 
The Legendre transform of $\ln Z_{gc}$ with respect to $\alpha_i$ is a function of $\beta$ and $N_{i,gc}$ defined by
$\ln \tilde Z_{gc}= \ln Z_{gc} -\sum_i \alpha_i N_{i,gc}$  and satisfies
\be
d \ln \tilde Z_{gc}= -E_{gc} d\beta - \sum_i   \alpha_i d N_{i,gc} \;.
\ee 
Thus we can alternatively calculate the grand canonical entropy by integrating $-E_{gc}(\beta,N_{i,gc})$ with respect to $\beta$ at fixed $N_{i,gc}$ to determine $\ln \tilde Z_{gc}$ and then find the entropy from $S_{gc}(\beta,N_{i,gc}) = \ln \tilde Z_{gc} + \beta E_{gc}$.   

 The state density $\rho(E,N_p,N_n)$ is related to the partition function $Z_{gc}(\beta,\alpha_p,\alpha _n)$ by
the three-dimensional inverse Laplace transform 
\begin{eqnarray}
\label{3DiL}
\rho(E,N_p,N_n) = {1\over (2 \pi i)^3}
\int^{i\infty}_{-i\infty} d \beta\, \int_{-i\infty}^{i\infty} d \alpha_p \int^{i\infty}_{-i\infty}  \,d \alpha_n\,  \nonumber \\ \times e^{
\beta E -\alpha_p N_p - \alpha_n N_n} Z_{gc} (\beta,\alpha_p,\alpha_n) \;.
\end{eqnarray}
Normally one carries out the integration over all three variables in the three-dimensional (3-D)
saddle-point approximation, resulting in the formula for the state density~\cite{BM}
\be
\label{r3D}
\rho(E,N_p,N_n) = {1\over(2 \pi)^{3/2}}
\left|{\partial (E,N_p,N_n) \over 
\partial (\beta,\alpha_p, \alpha_n)}\right|^{-1/2}e^{S_{gc}},
\ee
where the values of $\beta,\alpha_p,\alpha_n$ are determined from $E,N_p,N_n$ by the saddle-point conditions
\begin{eqnarray}
\label{3D-saddle}
 E &= - {\partial \ln Z_{gc} \over \partial \beta}= E_{gc}(\beta,\alpha_p,\alpha_n) \;, \nonumber \\
 N_i & = {\partial \ln Z_{gc} \over \partial \alpha_i} = N_{i,gc}(\beta,\alpha_p,\alpha_n) \;.
\end{eqnarray}

Using Eqs.~(\ref{3D-saddle}), the Jacobian in (\ref{r3D}) can be written as the determinant of the matrix of second
derivatives of the logarithm of the grand canonical partition function with respect to $\beta,\alpha_p,\alpha_n$.  

\subsection{Ensemble reduction}
\label{saddle-app}

To compare the mean-field entropies to the canonical SMMC entropies we need to reduce the grand canonical
ensemble of the mean-field formalism to the canonical ensemble.   For
approximate theories such as the HF and HFB, the only consistent way (in the
sense of Appendix II) to carry out the reduction is the
variation-after-projection method (VAP), but this is difficult andm as far as
we know, has never been put into practice for calculating level densities in
heavy nuclei.  We will therefore only consider simpler reduction methods,
recognizing that they cannot be free from ambiguity.  

A straightforward way to determine a canonical entropy
is to separate the 3-D saddle-point integration Eq.~(\ref{3DiL}) into two steps,
integrating first over the chemical variables $(\alpha_p,\alpha_n)$.  
This yields the following expression for the integrand of the $\beta$
integration:
\be
\label{2Dsp}
\zeta^{-1}Z_{gc}(\beta,\alpha_p,\alpha_n) e^{\beta E-\sum_i \alpha_i N_i} \;,
\ee
where 
\be
\label{zetasp}
\zeta = 2 \pi \left|{\partial (N_p,N_n) \over 
\partial (\alpha_p, \alpha_n)}\right|^{1/2}\;.
\ee
and $\alpha_p,\alpha_n$ are determined by the 2-D saddle-point conditions $N_i = \partial \ln Z_{gc} /\partial \alpha_i$ ($i=p,n$).
Comparing with the integrand in Eq.~(\ref{s2r}), we can identify the approximate 
canonical partition function as
\be
\label{Zgc2c}
\ln Z_c \approx \ln Z_{gc} -\sum_i \alpha_i N_i -\ln \zeta \;.
\ee
If we carry out in a second step the $\beta$ integration of Eq.~(\ref{2Dsp}), we obtain an expression of the same form 
as in Eq.~(\ref{s2r}) where the approximate canonical entropy is given by
\be
\label{Sgc2c}
S_c \approx S_{gc} - \ln \zeta \;.
\ee
The expression we find for the state density $\rho$ is equivalent to Eq.~(\ref{r3D}) of the 3-D saddle-point approximation~\cite{det}.

However, the above result does not take into account
the variation of the prefactor $\zeta$ with respect to $\beta$.
If we consider this dependence explicitly when we perform the $\beta$ integration, the saddle-point condition that determines $\beta$ in terms of the energy $E$ becomes $E=E_\zeta$, 
where $E_\zeta$ is an approximate canonical energy given by
\be \label{E0}
E_\zeta= E_{gc} - \delta E,\,\,\,{\rm with}\,\,\,\delta E =-{d \ln\zeta\over d \beta} \;.
\ee
The level density is then given by the canonical form (\ref{s2r}), 
where the canonical entropy is
\be
\label{S_c} 
S_{c}(\beta,N_p,N_n) = S_{gc} - \ln \zeta +\beta \delta E \;.
\ee
A simple model is presented in Appendix I, showing that the
saddle-point shift $\delta E$ in Eq.~(\ref{E0}) improves the accuracy of the 
calculated $S_c(\beta)$ and $\rho(E)$.

\subsection{Discrete Gaussian model and the particle-number fluctuation}
\label{discrete-gaussian} 

Another source of error may arise from the Gaussian approximation in the 2-D saddle-point
integration used to derive Eq.~(\ref{zetasp}).  For example, suppose that the grand
canonical partition function were dominant by a single nuclide $(N_p,N_n)$ in
some range of $\alpha_p,\alpha_n$.   Then we could approximate 
$Z_{gc}(\beta,\alpha_p,\alpha_n) \approx Z_c(\beta,N_p,N_n) \exp(\alpha_p N_p +
\alpha_n N_n)$.  Treating this in the saddle-point Gaussian integration gives $\zeta=0$ (since $N_i$ are independent of $\alpha_j$),
rather than the correct answer of $\zeta=1$.  The problem can be repaired by 
recognizing that $N_p$ and $N_n$ are discrete integers, not continuous variables. 
We calculate the matrix $\partial N_i / \partial \alpha_j$ as before but now we
calculate $\zeta$ as a discrete sum over particle numbers $N_i$,
\be\label{zeta-2} 
\zeta = \sum_{N'_i,N'_j} \exp\left(-\frac{1}{2} \sum_{i,j} {\partial N \over \partial
\alpha}\bigg |_{ij}^{-1} (N'_i - N_i)(N'_j - N_j) \right) \;. 
\ee  
Expression (\ref{zeta-2}) for $\zeta$ reduces to the saddle-point result (\ref{zetasp}) in the limit when the r.h.s.~of (\ref{zetasp}) is large, but has the advantage that it is always larger than $1$ and it approaches $1$ when the r.h.s.~of (\ref{zetasp}) goes to $0$.

 To see more physically how the approximation (\ref{zeta-2}) works, consider the case
when there is only one type of particles, say neutrons, and the Hamiltonian used in the Gibbs density operator is independent of $\beta$ and $\alpha_i$.  The required derivative is then
\be
\label{deltaN2}
{\partial^2 \ln Z_{gc}\over\partial  \alpha_n^2} = {\partial N_n \over
\partial \alpha_n} = \langle (\Delta N_n)^2\rangle \;,
\ee
where $ \langle (\Delta N_n)^2 \rangle = \langle \hat N_n^2 \rangle - N_n^2$
is the neutron-number 
fluctuation in the grand canonical ensemble. Carrying out the Gaussian
saddle-point integration,  $\zeta^{-1}$ in the 2-D saddle-point approximation (\ref{2Dsp}) becomes 
\be
\label{zeta-s}
\zeta_n^{-1} = \left(2 \pi\langle (\Delta N_n)^2\rangle\right)^{-1/2}\;.
\ee
$\zeta_n^{-1}$ is just the ratio of
states with particle number $N_n$ to the total number of states in an ensemble
in which the particle number $N'_n$ is distributed as a discrete Gaussian 
\be
\label{gauss-P-N}
P_{N'_n} = \zeta_n^{-1} e^{-(N'_n-N_n)^2/2 \langle (\Delta N_n)^2\rangle}
\ee
in the limit that $\langle (\Delta N_n)^2\rangle >> 1$.  

The finite-temperature mean-field approximation also provides a many-particle density matrix, so that the
particle-number fluctuation can be calculated directly from this density matrix (see Eqs.~(\ref{DN^2}) and (\ref{Nop}) below). In fact, this direct calculation is much easier to carry out than calculating numerically
the matrix of second derivatives of the logarithm of the partition function.  However,
because the canonical reduction is not carried out in a variational way,
the two methods are not guaranteed to give the same answer.  In the
sections below, we will examine and compare both methods of carrying out the canonical reduction.

In the finite-temperature mean-field approximations we use here, the off-diagonal particle-number correlations $\langle \Delta N_i \Delta N_j\rangle$  vanish for $i\neq j$ and $\zeta$ factorizes into two separate factors for protons and neutrons
\be
\label{zeta}
\zeta = \zeta_p \zeta_n,\,\,\,{\rm where}\,\,\,
\zeta_i =  
\sum_{N'_i} e^{-(N'_i-N_i)^2/2  \langle (\Delta N_i)^2 \rangle}  \;.
\ee 
$\zeta_i$ in Eq.~(\ref{zeta}) can be considered as the partition function which describes the fluctuations in the number of particles of type $i$.  The reduction from the grand canonical to the canonical partition function
is then given by
\be
\label{factorization}
Z_{gc}(\beta,\alpha_p,\alpha_n) e^{-\sum_i \alpha_{i} N_i} \approx
Z_c(\beta,N_p,N_n)  \zeta_p \zeta_n \;.
\ee
Relation (\ref{factorization}) describes the factorization of the grand canonical partition function into a canonical partition function and particle-number fluctuation partition functions.  It is exact in the simple model presented in Appendix I.

In summary, the saddle-point approximation breaks down when 
$2\pi \langle (\Delta N_i)^2 \rangle \leq 1$.
However, $\zeta$ in the discrete Gaussian model (Eqs.~(\ref{zeta}) and (\ref{factorization}) or Eq.~(\ref{zeta-2})) always 
satisfies $\zeta \geq 1$ and can be used even when the 
particle-number fluctuation is small.  We will demonstrate the improvement to the saddle-point formula in this limit
in the case of \dy~ (Sec.~\ref{dy}) where pairing correlations are weak.
  
\subsection{Spin-parity projected level density} 
 
The ultimate goal is to calculate  the 
spin-parity projected densities $\rho_{J^\pi}(E)$, defined as the number of 
levels of given angular momentum $J$ and parity $\pi$ per unit energy, not 
counting the $2J+1$ magnetic degeneracy of the levels.  The spin-dependent 
level densities $\rho_{J^\pi}(E)$ can be calculated through an angular momentum projection.
 The present paper is mainly focussed on the total state density and we will
not examine in details spin-projection methods.  However,
to make at least a tentative comparison of the level density at the neutron resonance threshold we will
calculate them taking a simplified
model for the spin-parity projection.  We follow 
common practice and assume the spin distribution is Gaussian in the three 
components of the angular momentum vector $\vec{J}$~\cite{er60}.  
Then the fraction of 
levels having angular momentum $J$ 
is given by 
\be 
\label{PJ} 
P_J \approx  
\sqrt{1\over  2\pi}  {J+\tfrac{1}{2}\over \sigma^3} 
\exp\left(-{(J+\tfrac{1}{2})^2\over 2\sigma^2}\right) \;. 
\ee 
Here a pre-exponential factor of $(J+\frac{1}{2})^2$, arising from the
three-dimensional volume 
element of $\vec{J}$, is reduced to the first power of $(J+\frac{1}{2})$ by
dividing out the $( 2J +1)$ magnetic degeneracy factor.
 The parameter $\sigma$, known as the spin cutoff 
parameter, is estimated from the second moment of $J_z$ 
\be\label{sigma2} 
\sigma^2 = \langle J_z^2 \rangle \;. 
\ee 
The normalization condition of $P_J$ in (\ref{PJ}) is  $\sum_J (2J+1)P_J=1$. 
Assuming equal positive- and negative-parity level densities (usually 
justified at the neutron binding energy), we have 
\be 
\label{spectro} 
\rho_{J^\pi}(E) \approx {1\over 2} P_J \rho(E) \;. 
\ee 
 
\section{SMMC results}\label{SMMC}
 
The SMMC method is formulated in the framework of a 
CI spherical shell-model Hamiltonian.  The CI 
Hamiltonians shown to be amenable to Monte-Carlo sampling contain one- plus 
two-body operators, with the two-body part restricted to interactions that 
have a ``good sign" in the grand canonical formulation~\cite{sign}. In finite nuclei, 
the method is implemented with particle-number projection~\cite{or94} for 
both protons and neutrons, and the calculated observables are the 
expectation values in the canonical density matrix.
In particular, we consider here the 
nuclei \sm~and \dy. The nucleus \sm~is an example of a heavy spherical 
nucleus whose ground state has significant correlation energy associated 
with pairing, while the nucleus \dy~ has a deformed ground state with weak 
pairing. 
 
The parameterization of the Hamiltonian and other aspects of the  
SMMC calculation have been published elsewhere~\cite{al08,oz13}.  For 
reference, Table I and its caption describes the model space employed in the calculations. 
 \begin{table}[htb]  
\begin{center}  
\begin{tabular}{|c|cccc|}  
\hline  
 &  $N_i$  &  $d_i $ & $S_{c,i}(0)$ & $S_{gc,i}(0)$\\ 
\hline  
\sm~n & 16 & $8.5 \cdot 10^{14}$ & 34.38 & 36.55\\ 
\sm~p & 12& $5.6 \cdot 10^{9}$   & 22.44 & 24.43\\ 
\dy~n & 26& $1.6 \cdot 10^{18}$ & 41.95 & 44.25\\ 
\dy~p &16 & $6.3 \cdot 10^{10}$  & 24.86 & 26.92\\ 
\hline  
\end{tabular}  
\caption{Model space parameters for the SMMC Hamiltonian in \dy~and \sm~ 
and corresponding canonical and grand canonical entropies at $\beta=0$. The single-particle basis employed
in the CI spherical shell model consists of the orbitals
$0g_{7/2},1d_{5/2},1d_{3/2,}2s_{1/2},0h_{11/2},1f_{7/2}$ for protons and
$0h_{11/2},0h_{9/2},1f_{7/2},1f_{5/2},2p_{3/2},2p_{1/2}, 0i_{13/2},1g_{9/2}$ for
neutrons. The numbers of the single-particle states (including their
magnetic degeneracy) are $D_p = 40$ and $D_n = 66$.  $N_i$ are the
number of valence particles of type $i$ (where the index $i$ distinguishes
neutrons and protons), and $d_i$ is the dimension of the many-particle model
space for particles of type $i$.  The canonical and grand canonical
entropies $S_{c,i}(0)$ and $S_{gc,i}(0)$ are calculated from
Eqs.~(\ref{Sc0}) and (\ref{Sgci0}), respectively.  
}
\label{d}  
\end{center}  
\end{table}  
 
The canonical energy $E_c = \langle \hat H \rangle_{N_p,N_n}$ and the  
mean square angular momentum $\langle \hat J^2 \rangle_{N_p,N_n}$ at fixed numbers of protons and neutrons are calculated directly in SMMC as a function of $\beta$. 
Table \ref{E-limits} shows the SMMC energies at $\beta=0$ (i.e., the infinite temperature limit) and 
at high $\beta$ extrapolated to infinity.
The energy at $\beta=0$  is largely determined by the one-body part of the Hamiltonian in the grand canonical ensemble.  There is no contribution from the direct  
component of the interaction but there is a small contribution from the exchange terms.  
\begin{table}[htb]  
\begin{center}  
\begin{tabular}{|c|c|ccc|cc|}  
\hline  
nucleus  &  $\beta$   &  SMMC & HF & HFB  &\multicolumn{2}{c|}{correlation energy}   \\ 
         &            &       &    &    &  HF/HFB  & \; missing \\ 
\hline  
\sm & 0 & -119.15 & -119.0 & -119.0 & &\\ 
 & $\infty$ & $-235.65 \pm 0.015$ & -230.69   & -232.51 & 1.82  & 3.14\\ 
\hline 
\dy & 0 & -238.35  & -238.12 & -238.12 && \\  
& $\infty$ &$-375.39 \pm 0.02$ & -371.78  & -371.91 & $ 11.41$  & 3.48 \\ 
\hline  
\end{tabular}  
\caption{Limiting values of the energies (MeV) calculated by SMMC, HF, and 
HFB for \sm~and \dy.  
The HF/HFB correlation energy is the energy difference between HF and 
HFB ground states for \sm, and the difference between spherical and
deformed ground states for \dy.  The term ``missing" denotes the 
differences between the HFB energies (with both pairing and deformation)
and the SMMC ground-state energies. The SMMC energies at $\beta = \infty$ include extrapolation and statistical sampling errors 
\cite{al08}.} 
\label{E-limits}  
\end{center}  
\end{table}  
 
\begin{figure}[htb!]  
\begin{center}  
\includegraphics[width= 9.1 cm]{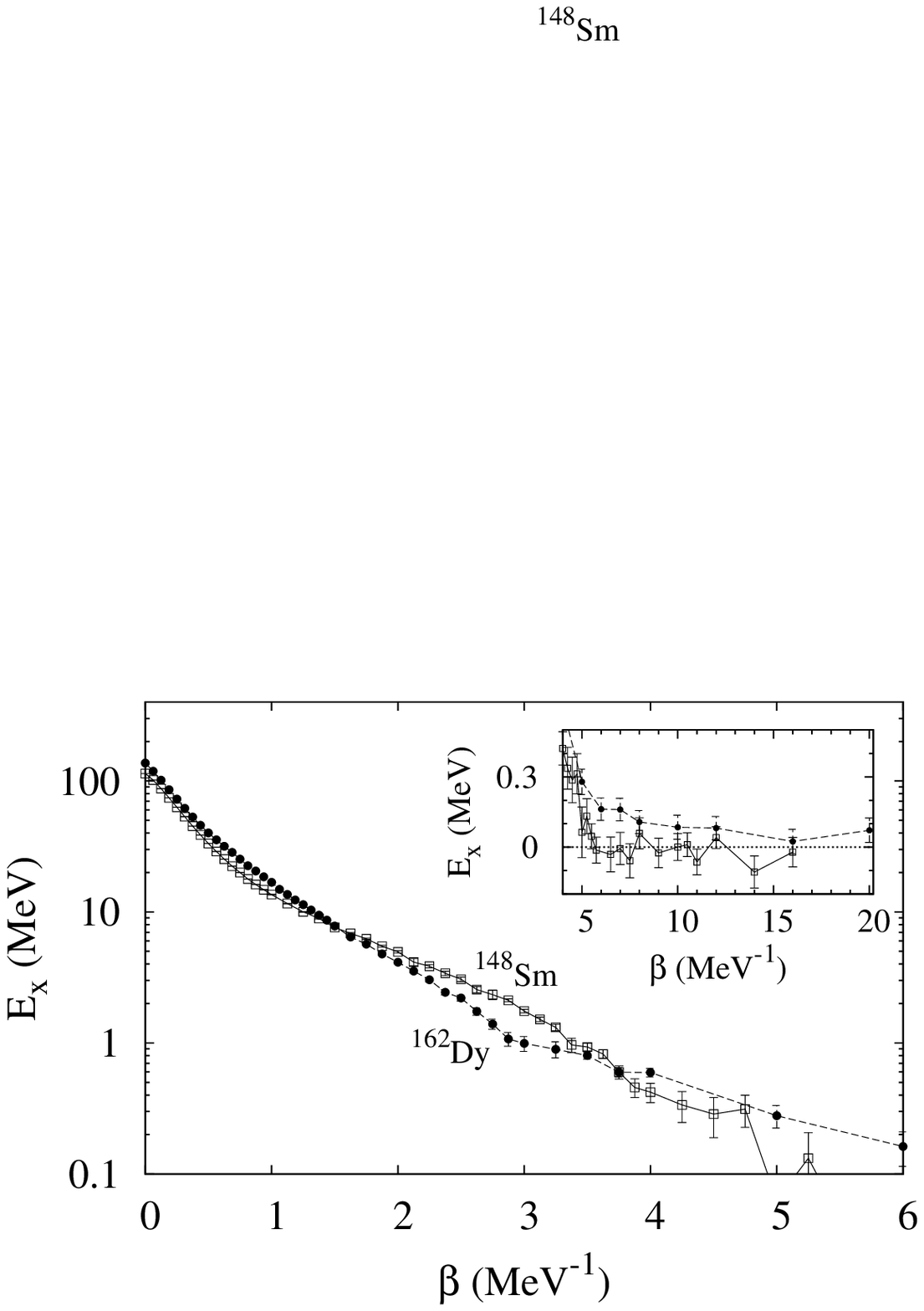}  
\caption{
Canonical excitation energies $E_x$  (on a logarithmic scale) versus $\beta$ calculated by the SMMC
for \sm~(open squares) and for \dy~(solid circles), with lines drawn to 
guide the eye. The inset shows the
large $\beta$ values using a linear scale for the excitation energy.  The
Monte Carlo statistical errors are about $0.1$ MeV or smaller.
} 
\label{e-smmc}  
\end{center}  
\end{figure}

The variation of the excitation energy $E$ with $\beta$ is shown in Fig.~\ref{e-smmc} using a logarithmic scale for the energy. 
The excitation energy of \dy~is higher than that of \sm~ from $\beta=0$ to 
$\beta\approx 1.5$ and is then lower up to $\beta\approx 3.5$. 
The higher excitation energy in \sm~near $\beta=3$ is likely due to the 
collapse of strong pairing in that nucleus.  Similarly, the higher \dy~excitation energy  at 
$\beta\approx 1$ may be ascribed to the loss of deformation energy 
in that temperature region.

\begin{figure}[htb]  
\begin{center}  
\includegraphics[width= 9.1 cm]{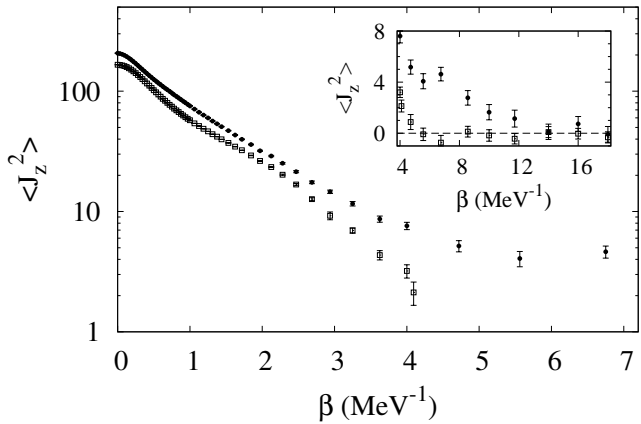}  
\caption{ SMMC values of $\langle J_z^2\rangle=\langle \vec J^2\rangle/3$ (shown using a logarithmic scale)  vs. $\beta$  in the canonical ensembles for \sm~ (open squares) and 
\dy~(solid circles). The inset shows $\langle J_z^2\rangle=\langle \vec J^2\rangle/3$ (using a linear scale) for larger $\beta$ values.} 
\label{Jsq}  
\end{center}  
\end{figure}  
\begin{figure}[htb]  
\begin{center}  
\includegraphics[width= 9.1 cm]{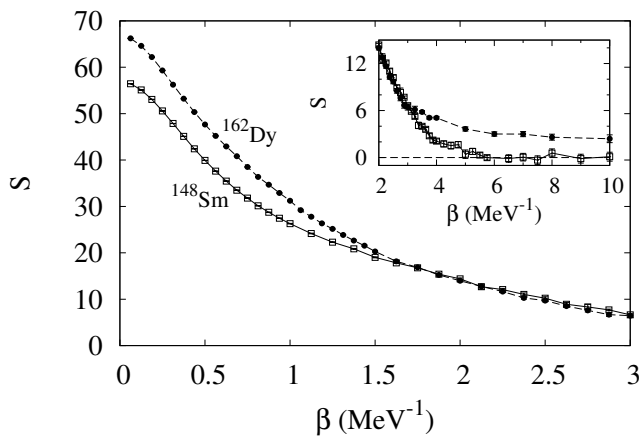}%
\caption{SMMC entropies of \sm~(open squares) and \dy~(solid circles) for $\beta < 3$ 
MeV$^{-1}$.  
In this range, the Monte Carlo errors are the size of the symbols or
smaller. The inset shows the entropies for the larger $\beta$ values.
} 
\label{figSa}  
\end{center}  
\end{figure}  
 
We also need the $\langle \hat J^2\rangle$ ensemble averages to calculate the spin-dependent level
densities.  These are shown in Fig.~\ref{Jsq}  for \sm~ and \dy.  
The higher  values of $\langle \hat J^2\rangle$ of \dy~ at high $\beta$ are largely  
due to its 
deformation and its low-lying first excited $2^+$ state at $\sim 0.08$ MeV.  In contrast,  $\langle \hat J^2\rangle$ for \sm~decreases dramatically 
at high $\beta$, as expected for a nucleus with a $J=0$ ground state 
and a gap of $\sim 0.5$ MeV to the first excited $J=2$ state. At low $\beta$, the remaining enhancement 
for \dy~is due to its larger number of active valence nucleons in the model 
space.  The errors shown in Fig.~\ref{Jsq} are statistical errors from the Monte Carlo 
sampling. 
 
We next apply Eq.~(\ref{S}) to compute the canonical SMMC entropy. 
We start from $\beta = 0$ with the 
initial value of the canonical entropy given by Eq.~(\ref{Sc0})  
and use the relation
\be 
\int_{E_c(\beta)}^{E_c(0)} \beta' d E_c= -\beta E_c(\beta) + \int_0^\beta E_c(\beta') d\beta' \;, 
\ee 
where $E_c(\beta)$ is the canonical thermal energy calculated in SMMC as the thermal expectation
 value of the Hamiltonian.  The results  are
shown in Fig.~\ref{figSa}, with the main figure showing the low to
intermediate values of $\beta$, and the inset showing the large values of
$\beta$.
 
The \sm~ and \dy~ entropies are nearly  
equal for $\beta$ in the range $1.5-3.0$ \Mevinv.  We do not know any obvious reason  
why that should be the case.  At higher values of $\beta$, shown in the inset, one observes
 a difference between the two nuclei.  For \sm~at $\beta\ge 6$ MeV$^{-1}$ the 
entropy is essentially zero, as expected at low temperatures for 
even-even nucleus with a pairing gap.  On the 
other hand, the entropy  of \dy~remains a couple of units higher than zero up 
to at least $\beta\sim 10$~\Mevinv.  This is due to the low excitations 
associated with the ground-state rotational band, together 
with the weak pairing in this nucleus. 
 
\begin{figure}[htb]
\begin{center} 
\includegraphics[width= 9.1 cm]{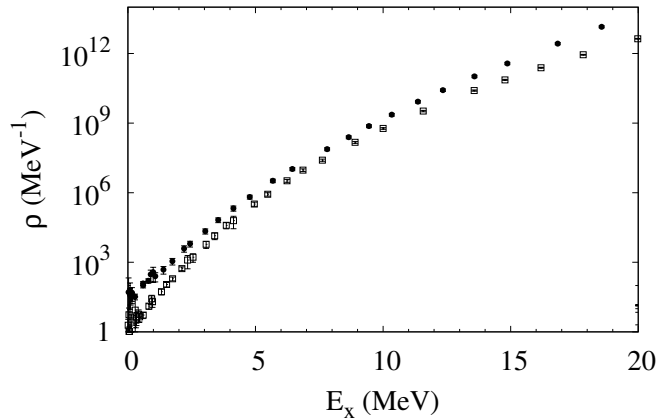}
\caption{ State densities vs.~excitation energy $E_x$ calculated in SMMC
using the saddle-point expression Eq.~(\ref{s2r})
for \dy~(solid circles) and \sm~(open squares).
The range is truncated at the lower end because the
saddle-point approximation breaks down when there are only a few
levels in a range $\beta^{-1}$ around $E$.  
} 
\label{rhos}  
\end{center}  
\end{figure}

The state densities calculated from Eq.~(\ref{s2r}) are shown in
Fig.~\ref{rhos}.  As expected, the state density is higher for \dy~
than \sm~ at low excitation energy.  Interestingly, they become 
much closer at excitation energies in the range 5-10 MeV.
As a check on the quality of the CI Hamiltonian, we compare the 
calculated level densities with the experimental $s$-wave resonance spacings 
$D = [\sum_J \rho_{J^\pi}(E)]^{-1}$, measured at an energy $E$ that corresponds to the neutron separation energy.  
These are calculated from the spin-parity dependent level density (\ref{spectro}) and (\ref{PJ}), taking the 
spin cutoff parameter $\sigma$ from Eq.~(\ref{sigma2}).   
The results are shown in Table III. 
The good agreement 
gives us confidence that the Hamiltonian is realistic enough to provide 
useful tests of the HF and HFB approximations. 
\begin{table}[htb]  
\begin{center}  
\begin{tabular}{|c|cc|cccc|}  
\hline  
Nucleus & $E_x$ (MeV)& $J^\pi$ &        \multicolumn{4}{c|}{$D$ (eV)}  \\ 
 
  &  &  &      SMMC    &  \;HF & HFB   &     Exp.     \\  
\hline  
\sm & 8.1 &$(3^-,4^-)$ & $3.7\!\pm\!0.6$  & &4.1  &  5.7 \\ 
\dy & 8.2 &$(2^+,3^+)$ & $2.4\!\pm\!0.3$  & 0.5 &   &      2.4 \\  
\hline  
\end{tabular}  
\caption{$s$-wave resonance spacings $D$ at the excitation energy $E_x$ 
that corresponds to the neutron binding energy. $J^\pi$ are the values of spin and parity of the relevant neutron resonance levels. The  SMMC, HF and HFB 
results are compared with the experimental values from 
Ref.~\onlinecite{RIPL}. The HF spacing is calculated using the model of Sec.~\ref{spacing} to estimate the contribution of rotational bands.
}  
\label{D}  
\end{center}  
\end{table}  
  
\section{The finite-temperature HF approximation} 
\label{HF} 
 
We first consider the FTHF approximation.  
It is derived by minimizing the grand potential $\Omega$ in terms 
of a trial uncorrelated many-particle  density matrix.  Such a density is uniquely characterized by 
its one-body density matrix $\varrho_{kl}$, where $k,l$ label single-particle  
orbitals in the model space.  

For simplicity, we consider only one type of particles and the results are easily generalized
 to both protons and neutrons. At the FTHF minimum, the one-body density $\varrho$ satisfies the 
self-consistent equation 
\be\label{rho} 
\varrho = { 1\over e^{\beta h_\varrho - \alpha} +1} \;,   
\ee 
where $h_\rho = t + v\varrho$ is the one-body HF Hamiltonian, expressed respectively in 
terms of the one- and two-body matrices $t$ and $v$ of the configuration-interaction shell model Hamiltonian.  The value of $\Omega$ at the HF minimum is given by   
 \be\label{Omega-HF} 
 \beta\Omega_{HF}= -\ln Z_{HF} = \beta E_{HF} - S_{HF} - \alpha N_{HF} \;, 
 \ee 
where $ Z_{HF}$ is the HF approximation to the grand canonical partition 
function.  The thermal HF energy $E_{HF}$ is calculated as
\be\label{E-HF} 
 E_{HF} = \tr (t\varrho) + \frac{1}{2} \tr(\varrho v\varrho)  \;,
\ee 
the entropy $S_{HF}$ is given by
 \begin{eqnarray}\label{S-HF} 
S_{HF} & = & -\tr(\varrho \ln\varrho) -\tr[(1-\varrho)\ln(1-\varrho)]  \nonumber \\
 &= &-\sum_\lambda f_k  \ln f_k -\sum_k 
(1-f_k )\ln(1-f_k ) \;, 
 \end{eqnarray} 
 and the average number of particles $N_{HF}$ is 
computed as
 \be\label{N-HF} 
 N_{HF} = \tr\, \varrho \;. 
 \ee 
The occupation probabilities 
$f_k=[1+e^{\beta(\epsilon_k-\mu)}]^{-1}$ are the usual Fermi-Dirac 
occupations where $\epsilon_k$ are the single-particle HF energies at temperature 
$T$.   
 
It is not obvious from Eqs.~(\ref{Omega-HF}), (\ref{E-HF}) and (\ref{N-HF}) that $\ln Z_{HF}$ satisfies the thermodynamic
derivative relations (\ref{3D-saddle}) due to the dependence of
the HF single-particle energies and density on $\alpha$ and $\beta$. 
Nevertheless, the contributions to the derivatives that originate in the implicit dependences
on $\alpha$ and $\beta$ vanish.  In particular, the quantities defined in
the HF theory satisfy
\be\label{E-N-HF1} 
  -  {\partial \ln Z_{HF} \over \partial \beta}= E_{HF} \;,\;\;\;
 {\partial \ln Z_{HF} \over \partial \alpha_i} = N_{i,HF}  \;. 
\ee 
The proof is provided in Appendix II. The validity of these relations in
FTHF guarantee that the grand canonical HF entropy $S_{HF}$ satisfies the relation
$ d S_{HF} = \beta d E_{HF}$ at fixed  average particle numbers $N_{i,HF}=N_i$, 
and thus we can compute the entropy in the
same way as we did in the SMMC.
 
\subsection{Approximate canonical projectors in FTHF} 
 
The canonical partition function can be approximated either in the
saddle-point approximation of Sect.~\ref{saddle-app} or in the discrete
Gaussian model of Sect.~\ref{discrete-gaussian}.  However, the connection
to particle-number fluctuations is more tenuous because the second derivative expressions for $\ln Z_{gc}$ in terms of particle-number fluctuations such as
Eq.~(\ref{deltaN2}) no longer holds for $\ln Z_{HF}$.  Nevertheless, it is interesting to compare $\zeta$ computed with the 
full matrix $\partial N_i/\partial \alpha_j$ to that computed from the
particle-number fluctuations. We make such a comparison for \dy~in the next section.

\subsection{Application to \dy}\label{dy}
 
Here we discuss the strongly deformed nucleus \dy.  The pairing in this 
nucleus is weak, so that the FTHF is the appropriate mean-field theory.   

\subsubsection{Number partition function $\zeta$}
 
\begin{figure}[htb]  
\begin{center}  
\includegraphics[width= 9.1 cm]{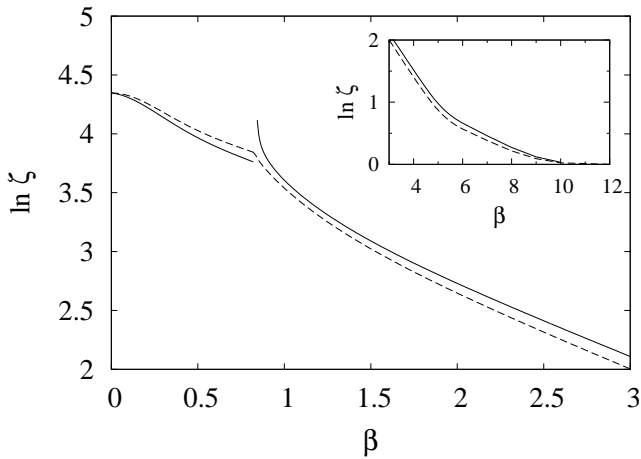} 
\caption{
$\ln \zeta$ vs.~$\beta$ in FTHF for \dy.  Solid line: $\ln \zeta$
calculated from Eq.~(\ref{zeta-2}). Dashed line: $\ln\zeta$ calculated from
Eq.~(\ref{zeta}).
}
\label{lnzeta_dy162}  
\end{center}  
\end{figure}

We first examine the number partition function $\zeta$ obtained from the  
approximations we presented in Sec.~II. The diagonal particle-number 
fluctuations in FTHF are given by
\be\label{DN^2} 
\langle (\Delta \hat N_i)^2\rangle = 
\tr [\varrho_i(1-\varrho_i)]  
\ee 
and the off-diagonal ones are zero.  The corresponding $\zeta$ obtained from 
Eq.~(\ref{zeta}) is shown as the dashed line in Fig.~\ref{lnzeta_dy162}. 
This is compared to $\zeta$ calculated from Eq.~(\ref{zeta-2}) (using the matrix $\partial N_i/\partial \alpha_j$), shown
as the solid line.  They differ by less than 10\%, except for a
tiny region near the spherical-to-deformed phase transition.
There the $\zeta$ calculated from the Jacobian of $\partial N_i/\partial \alpha_j$ diverges 
(when approached from the deformed side). Thus, it appears to be a very good approximation
to calculate $\zeta$ in term of the individual particle-number fluctuations as in Eq.~(\ref{zeta}).
In the next section, we shall see that this is not the case for the FTHFB
theory in the presence of strong pairing correlations.  In any case,
we will use Eq.~(\ref{zeta-2}) in the results shown below.

\subsubsection{Thermal excitation energy} 
 
The range of FTHF energies as a function of $\beta$ is shown in the fourth 
column of Table II. 
The high-temperature limit is very close to the SMMC value, 
since all correlation energies disappear in that limit.  At the 
other limit of large $\beta$, at or near the ground state, the SMMC energy is about  
3.5 MeV lower than its HF value.   
The HF ground-state energy has the 
correlations associated with the static deformation, but is 
missing the rotational energy and other correlation effects. 
It can also be seen from the fifth column of Table II that the HFB approximation 
hardly lowers the energy.  
\begin{figure}[htb]  
\begin{center}  
\includegraphics[width= 9.1 cm]{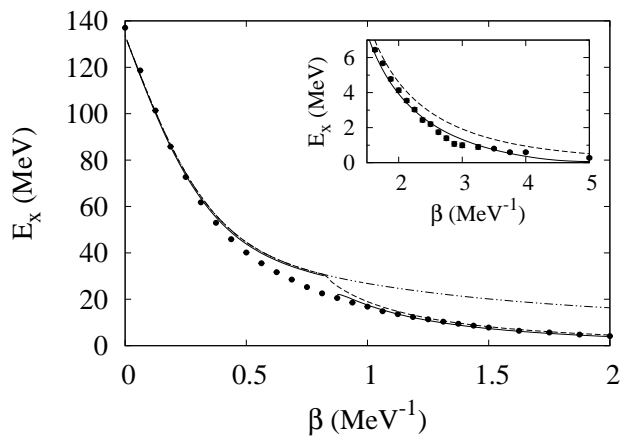} 
\caption{
The HF excitation energy of \dy~vs.~$\beta$.
The grand canonical HF energy (dashed line) is compared with the approximate
canonical energy $E_\zeta$ (solid line) from Eq.~(\ref{E0}).  The latter omits
the region near the shape transition point, where $\zeta$
becomes singular (see solid line in Fig.~\ref{lnzeta_dy162}).  We also show the
energy for the spherical HF solution (dashed-double dotted line) with
respect to the deformed ground-state energy.  The solid circles are the SMMC
excitation energies from Fig.~\ref{e-smmc}.  Inset: expanded energy scale
for higher $\beta$ values.}
\label{hf-dy162a}  
\end{center}  
\end{figure} 
The excitation energy $E_x(\beta)$ is shown in Fig.~\ref{hf-dy162a}, 
comparing its HF value (solid line) with the SMMC thermal energy (solid circles) from Fig.~\ref{e-smmc}.   
The HF density is spherical for $\beta < 0.9 $ and becomes deformed above 
that value.  The energy of the spherical HF solution at higher values of $\beta$ is 
shown in Fig.~\ref{hf-dy162a} by the dashed-double dotted line. 
We observe that the HF energy has a cusp at the 
onset of deformation. Extrapolating the energy of the spherical solution to large $\beta$, we find that the deformed ground-state solution is lower in energy than the spherical solution by 11.4 MeV.  
The SMMC energy, shown by the solid circles, is remarkably close 
to the HF energy in the region $0<\beta< 3$ MeV$^{-1}$ except in the  
vicinity of the shape phase transition. The cusp in the HF energy function disappears  
in the SMMC energy, leaving no trace of a shape phase transition.  The 
fact that mean-field theory over-emphasizes phase transitions in finite 
systems is well-known in nuclear theory; see, e.g., in Refs.~\onlinecite{es93,ro98,ka06,ga13}. 
The inset in Fig.~\ref{hf-dy162a} shows  
$E(\beta)$  at higher values of $\beta$ using a finer energy scale. We observe 
that at large $\beta$ the grand canonical HF energy (dashed line) overestimates the excitation energy but that the approximate canonical energy $E_\zeta$ of Eq.~(\ref{E0}) (solid line) is in overall good agreement with the SMMC excitation energy. 
 
\subsubsection{Entropies} 
 
\begin{figure}[htb]  
\begin{center}  
\includegraphics[width= 9.1 cm]{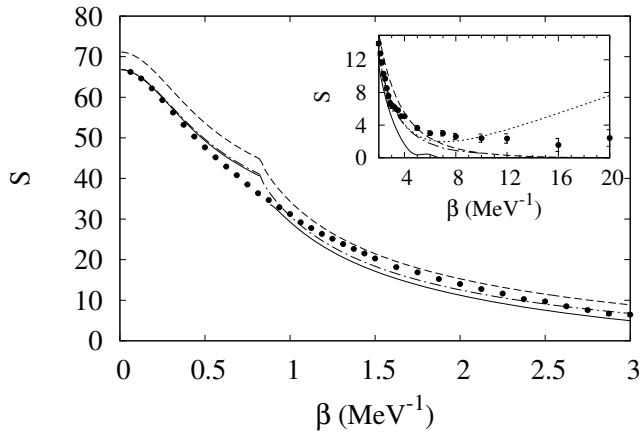}%
\caption{
Entropy of \dy~in the FTHF approximation, comparing the grand canonical
entropy (dashed line) with the canonical entropy defined in Eqs.~(\ref{S_c}) and (\ref{E0}) 
(solid
line). The dashed-dotted line is the approximate canonical entropy (\ref{Sgc2c}) in
the 3-D saddle-point approximation, i.e., without the correction 
term in Eq.~(\ref{E0}).  
The inset shows the large $\beta$ value.  The
calculations use the discrete Gaussian model formula (\ref{zeta-2}) for
$\zeta$.  The dotted line in the inset uses the saddle-point expression
Eq.~(\ref{zetasp}) for $\beta > 5$~\Mevinv.
} 
\label{gc2c}  
\end{center}  
\end{figure}

The grand canonical HF entropy $S_{HF}$  is shown in Fig.~\ref{gc2c} as a 
function of $\beta$ (dashed line) and compared with the approximate 
canonical entropy (\ref{S_c}) with $\zeta$ from the discrete
Gaussian formula Eq. (\ref{zeta-2}) (solid line).   
At $\beta=0$, the 
grand canonical HF entropy is larger than the canonical entropy due to
particle-number fluctuations.  The 
entropies at large values of $\beta$ are shown in the inset. The 
grand canonical HF entropy vanishes in the limit $\beta \to 
\infty$, as expected. 
However, the saddle-point canonical entropy calculated from Eqs.
(\ref{2Dsp}) and (\ref{zetasp}) increases at large values of $\beta$ (dotted
line in the inset), indicating the breakdown of the saddle-point
approximation to the particle-number projection.  
In contrast, the discrete Gaussian treatment (\ref{zeta-2}) gives
an entropy that approaches zero at high $\beta$, thus satisfying the
sum rule Eq. (\ref{sr}). For moderate and small values of $\beta$,
the entropy (\ref{S_c}) of the discrete 
Gaussian model (\ref{zeta-2}) essentially coincides with the saddle-point canonical entropy.
As noted earlier, the SMMC entropy remains nonzero
in the range $8 < \beta < 15$.  
We examine this further in the next paragraph.
\begin{figure}[h!]  
\begin{center}  
\includegraphics[width= 9.1 cm]{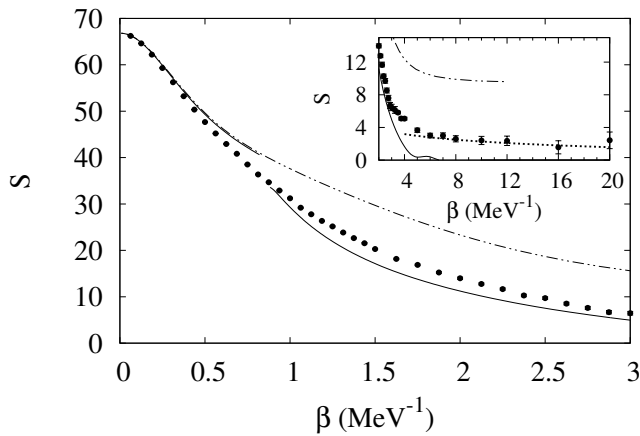}%
\caption{Approximate canonical HF entropy defined by Eqs.~(\ref{S_c}) and (\ref{E0}) for \dy~(solid line) is compared with the SMMC 
entropy (solid circles).  The dashed-double dotted line is the canonical
entropy of the spherical HF solution in the same approximation. The inset shows the entropies at large
values of $\beta$. The dotted line in the inset is the ground-state
rotational band contribution (\ref{S-rot}).
} 
\label{svb-dy162}  
\end{center}  
\end{figure}  

To compare the projected HF and SMMC entropies in more detail, we have replotted
them in Fig.~\ref{svb-dy162} with some additional information.
Both curves start at the same value at $\beta=0$ because the 
model spaces are identical. In the limit of large $\beta$, the  
SMMC entropy does not approach zero as fast as the canonical HF entropy. 
This is because the SMMC entropy includes a contribution from the ground-state 
rotational band, most visible at $\beta >5$~\Mevinv.  We can estimate this contribution 
as follows. The moment of inertia ${\cal I}_{gs}$ of the ground-state band 
of \dy~was determined to be ${\cal I}_{gs}/\hbar^2 = 35.8 \pm 
1.5$~\Mevinv~by fitting the low-temperature SMMC values of $\langle {\vec 
J}^2\rangle$ to $\langle {\vec J}^2\rangle= 2({\cal 
I}_{gs}/\hbar^2)T$~\cite{al08}.  For $\beta < 20$~\Mevinv, $T > \hbar^2/2 
{\cal I}_{gs}$, and we can treat the rotational motion classically. The 
classical partition function of the rotational band $J=0,2,\ldots$ is given 
by $Z_{rot} = {\cal I}_{gs} T/\hbar^2$. We can then calculate its entropy 
from $S_{rot}=- \partial F_{rot}/\partial T$, where $F_{rot}=-T \ln Z_{rot}$ 
is the free energy of the ground-state rotational band. We find 
\be\label{S-rot} 
S_{rot} = 1 + \ln\left(\frac{ {\cal I}_{gs}}{\hbar^2} T\right) \;.
\ee  
This contribution is described by the dotted line in the inset of 
Fig.~\ref{svb-dy162}, and is in good agreement with the SMMC entropy at 
large $\beta$ with no adjustable parameters. 

We also show in Fig.~\ref{svb-dy162}  the canonical entropy of the spherical 
HF solution (dashed-double dotted line).  This entropy approaches a finite non-zero value 
in the limit $\beta \to \infty$ (see inset) because of the large degeneracy 
of the spherical HF solutions at $T=0$.  There are 2 valence protons in the 
$0h_{11/2}$ orbitals and 6 valence neutrons in the $0h_{9/2}$ orbitals, 
leading to a highly degenerate ground state with a canonical entropy of $\ln 
\left[{12 \choose 2}{10 \choose 6}\right]=9.54$. The grand canonical HF 
entropy in this limit is even larger.  It can be calculated assuming the uniform 
filling of the valence degenerate orbitals in the $T\to 0$ limit of the HF 
approximation. The corresponding formula has the form of Eq.~(\ref{Sgci0}), 
where $\Omega_p=12$, $f_p=2/12$ and $\Omega_n=10$, $f_n=6/10$ and gives a 
grand canonical entropy of $13.33$. 
 
\subsubsection{Angular momentum fluctuations} 
  
\begin{figure}[bth]  
\begin{center}  
\includegraphics[width= 9.1 cm]{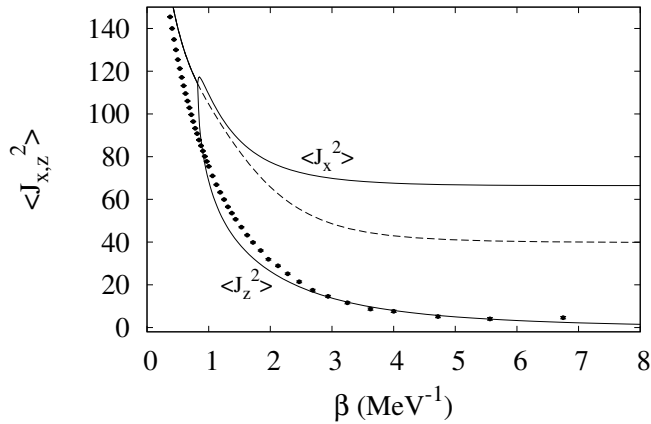} 
\caption{Second moments of the angular momentum in \dy.  
The solid lines are the HF results which exhibit a kink at the shape
transition point. The dashed line describes the spherical HF solution for
temperatures where the lowest equilibrium solution is deformed. These HF
moments may be compared with the SMMC moments shown by solid circles.  The
SMMC moments satisfy $\langle J_{x,y}^2 \rangle=\langle J_z^2\rangle=\langle
\vec J^2\rangle/3$.
} 
\label{J2_dy162}   
\end{center}  
\end{figure}

In HF, the variance of the angular momentum components $J_q$ ($q=x,y,z$)
can be calculated using Wick's theorem as
\be\label{DJ-HF} 
\langle (\Delta J_q)^2\rangle = \langle \hat J_q^2 \rangle -\langle \hat 
J_q\rangle^2 = \tr [j_q \,(1-\varrho) j_q\, \varrho] \;, 
\ee 
where $j_q$ is the matrix representing $\hat J_q$ in the single-particle space.  
When the HF equilibrium ensemble is axially symmetric around the $z$-axis,
$\varrho$ is invariant under rotations around the $z$ axis and $\langle
\hat J_x\rangle = \langle \hat J_y\rangle=0$. Assuming time-reversal invariance, we
also have $\langle \hat J_z\rangle=0$. It follows that the variances of the
angular momentum components are the same as the mean square moments.  In
 Fig.~\ref{J2_dy162}, we compare the HF mean square moments of $\hat J_x$ and
 $\hat J_z$ (solid lines) with the SMMC moments $\langle \hat J_x^2\rangle
= \langle
 \hat J_y^2\rangle= \langle \hat J_z^2\rangle =\langle \hat J^2\rangle/3$ (solid
 circles) in \dy~. The HF mean square moments of $ \hat J_x$ and $ \hat J_z$ coincide
 above the shape transition temperature, where the HF solution is spherical. 
 However, at large values $\beta$, the HF mean square moment of $ \hat J_x$ is
 much larger than the respective moment of $ \hat J_z$.  Since the deformed
 intrinsic ground state has good $K=0$, $\langle \hat J_z^2\rangle$
 approaches zero in the limit $\beta \to \infty$, while $\langle \hat J_x^2\rangle$
 remains finite and large in this limit.
We also show by dashed line $\langle \hat J_x^2\rangle$
 for the spherical HF solution.
 
\subsubsection{State density and level spacing}\label{spacing}

\begin{figure}[h!]  
\begin{center}  
\includegraphics[width= 9.1 cm]{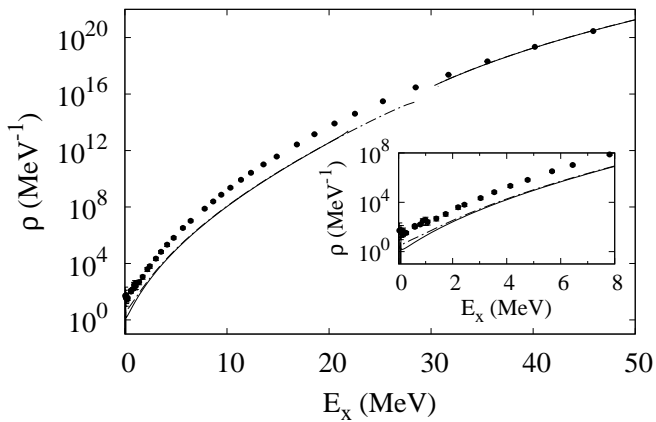} 
\caption{
The HF density of \dy~calculated in the saddle-point approximation (\ref{s2r}) using Eqs.~(\ref{E0}) and (\ref{S_c}) for the approximate energy and canonical entropy (solid line) is
compared with the SMMC state density (solid circles) as a function of
excitation energy $E_x$.  The gap in excitation energy reflects the discontinuity of the energy at the shape transition as is seen in Fig.~\ref{hf-dy162a}. The dashed-dotted line is the approximation
in which the $\delta E$ correction is neglected in the saddle-point energy Eq.~(\ref{E0}) and in the approximate canonical entropy of Eq.~(\ref{S_c}).  The inset shows an expanded scale at low excitation energies.
} 
\label{rho_dy162}  
\end{center}  
\end{figure}  
 
In Fig.~\ref{rho_dy162} we show the HF density vs.~$E_x$ in the saddle-point 
approximation (\ref{s2r}) (solid line) (where the approximate canonical entropy and energy include the $\delta E$ correction in Eqs.~(\ref{S_c}) and (\ref{E0}) respectively),  and
compare it with the SMMC state density (solid circles).
The kink in the HF density at $E \approx 31$ MeV signifies the shape 
transition from a deformed to spherical shape. At lower excitation energies, 
the HF state density underestimates the SMMC values; the SMMC density 
includes a contribution from rotational bands that are built on top of 
intrinsic $K$ states, and are not captured in the HF approximation.  Above 
the shape transition energy, the equilibrium shape is spherical and no 
longer supports rotational bands. The HF density is then very close to the 
SMMC density. 
  
We can try to repair the HF approximation by recognizing that the each of the 
deformed HF configurations represents a band~\cite{bj74}.  The angular momentum $J_z$ 
corresponds to the $K$-quantum number of the band. Assuming that $K$ is
Gaussian distributed, the $K$-dependent HF state density $\rho_K$ can be expressed
\be 
\rho_K(E) = P_K \rho_{HF}(E). 
\ee 
where
\be 
P_K = {1\over (2 \pi \sigma^2)^{1/2}} \exp\left(-{K^2\over 2 
\sigma^2}\right). 
\ee 
and 
\be
\sigma^2 = \langle J_z^2\rangle.
\ee
For each positive $K$ there will be a rotational band with
$J = K,K+1,...$.  For $K=0$ the sequence may skip odd or
even $J$ values.  For a complete treatment of the band
model one next introduces a moment of inertia of the
band to calculate the $J$-dependent level density.  However,
we do not wish to introduce new parameters that take us
beyond the HF theory so we assume that all the levels in
a band are degenerate.  The level density is then 
\be 
\rho_{J^\pi}(E) \approx {1\over 2}\sum_{K = 0}^J  \rho_{K}
(E),      
\ee 
treating the $K=0$ bands the same as the others.  The factor
of $1/2$ is for parity projection.
The resulting average resonance spacing at the neutron threshold of $E=8.2$ MeV for \dy~ is reported 
in Table III.  It underestimates the SMMC value by a factor of $\sim 5$.
This is a substantial disagreement; in our view uncovers a serious
problem with the HF theory of level densities in deformed nuclei.
 
\subsubsection{Independent-particle model} 
 
A common approximation in the calculation of state densities is to take
the HF ground-state mean field, and assume the excited states can be calculated
in the independent-particle model (IPM) with single-particle energies
given by that potential field.
For an axially deformed nucleus such as \dy,~the single-particle HF levels
come in doubly degenerate time-reversed pairs and for an even number of
particles the ground state is non-degenerate, so that the $T=0$ entropy is
zero. It is rather easy to carry out the exact paritcle-number projection in the IPM~\cite{be14}, so we will use it in the comparison.
\begin{figure}[h!]  
\begin{center}  
\includegraphics[width= 9.1 cm]{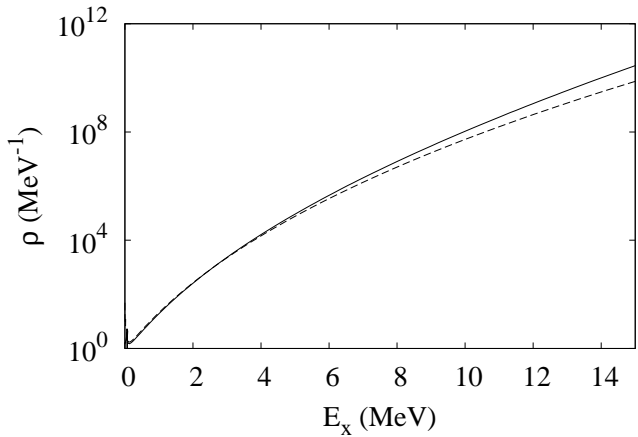} 
\caption{
The particle-projected IPM density of \dy~(dashed line) based on the zero temperature HF single-particle levels is compared with the HF density (solid
line from Fig. \ref{rho_dy162}) as a function of excitation energy $E_x$.  }
\label{rho_dy162_ipm}  
\end{center}  
\end{figure}  
 In Fig.~\ref{rho_dy162_ipm}, we compare the IPM state density with the
FTHF density determined by using Eq. (\ref{S_c}) for $S_c$.
They agree very well at low excitation energy.
At the neutron separation energy, $S_n\approx 8.2$ MeV,
the IPM state density is lower than the FTHF density by less than a factor of 2.  However, the
discrepancy increases with excitation energy, exceeding one unit entropy beyond ~15 MeV
excitation energy. At the shape phase transition, the IPM underestimates
the HF density by more than two orders of magnitude.
We conclude that IPM is a good
approximation at excitation energies that are small compared to the
shape transition energy, but not near and above this transition energy.

\section{The finite-temperature HFB approximation}\label{HFB}
 
The HFB is the preferred mean-field approximation for nuclei exhibiting 
strong pairing correlations.  Like the FTHF, the FTHFB is based on 
the grand canonical ensemble. However, unlike the FTHF, the simple approximate 
particle-number projection onto a canonical ensemble is not expected to be a 
good approximation at low temperatures. This will be evident as we go 
through the steps to calculate the level density of \sm, starting from the 
HFB energy function $E_{HFB}(\beta)$.  To simplify the notation, we consider only one type of particles as we did in the FTHF. The HFB thermal energy is expressed in terms of the  
normal and anomalous densities $\varrho,\kappa$ as
\be 
\label{EHFB}
 E_{HFB} = \tr (t\varrho) + \frac{1}{2} \tr(\varrho v\varrho) + \frac{1}{4} \tr(\kappa^\dagger v\kappa) 
 \;. 
\ee 
The HFB entropy $S_{HFB}(\beta)$ is given by 
\be\label{entropy-HF2} 
S_{HFB} =-\sum_k f_k \ln f_k -\sum_k (1-f_k)\ln(1-f_k) \;, 
 \ee 
where 
\be 
 f_k={1 \over 1+ e^{\beta E_k} } 
\ee 
are the quasi-particle occupations expressed in terms of the quasi-particle energies $E_k$.  

The HFB partition function satisfies relations similar to Eqs.~(\ref{E-N-HF1}), and thus the entropy can be computed from the energy function by an
integral similar to Eq.~(\ref{S}).  
All the expressions regarding the 
level density in Sec.~\ref{HF} carry over to the HFB approximation, 
except that the particle-number variance in Eq.~(\ref{DN^2}) is now 
calculated using all three Wick contraction terms in the expectation value 
$\langle a^\dagger_k a_k a^\dagger_l a_l\rangle$. This leads to an 
additional contribution from the anomalous density $\kappa$ 
\be 
\label{Nop} 
\langle (\Delta \hat N)^2\rangle = 
\sum_k \varrho_{kk} -\sum_{kl} \varrho_{kl}^2 + \sum_{kl}|\kappa_{kl}|^2 \;. 
\ee 
 
\subsection{Application to \sm}\label{sm}
 
   The mean-field ground state of \sm~is spherical and has a substantial
pairing condensate.  Thus FTHFB is the proper mean-field theory for this
nucleus. The pairing correlation energy, defined as the difference between
the HF and HFB ground-state energies is 1.82 MeV (see Table II).  Also
shown in Table II is the correlation energy of the SMMC ground state with
respect to the HFB solution (i.e., the difference between the HFB and SMMC ground-state
energies). This correlation energy is labeled ``missing'' in the table and is about $3$ MeV for \sm. 
The pairing transitions for protons and neutrons occur at $\beta= 2.1$
MeV$^{-1}$ ($E_x=5.9$ MeV) and $\beta=2.7$ MeV$^{-1}$ ($E_x=3.4$ MeV),
respectively.
 
\begin{figure}[htb]  
\begin{center}  
\includegraphics[width= 9.1 cm]{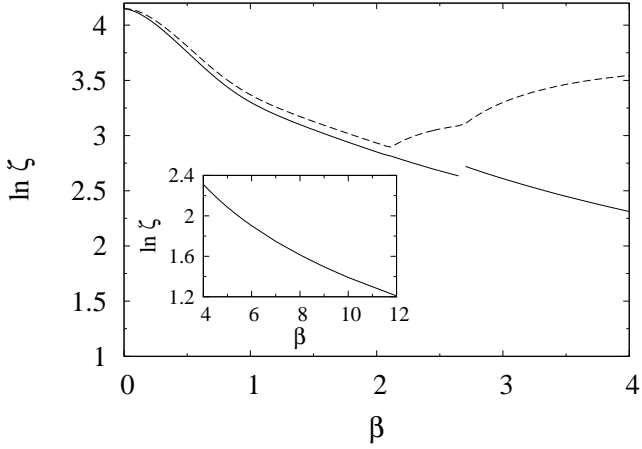} 
\caption{
$\ln \zeta$ vs.~$\beta$ in FTHFB for \sm.  Lines are as in
Fig.~\ref{lnzeta_dy162}. }
\label{lnzeta_sm148}  
\end{center}  
\end{figure}

We first examine $\zeta$, the factor used to convert the grand
canonical quantities to canonical in the HFB.  Naively, we may try
using the HFB particle-number fluctuation in Eq.~(\ref{Nop}) in Eq.~(\ref{zeta}).  The resulting $\zeta$ is shown as the dashed line in 
Fig.~\ref{lnzeta_sm148}.  This approximation is bound to fail at
low temperatures because the HFB ground state has a nonphysical
particle-number fluctuation.  We have also calculated $\zeta$ from Eq.~(\ref{zeta-2}) 
using the full $\partial N_i/\partial \alpha_j$ matrix suggested by the 3-D saddle-point
approximation.  This gives a better result for $\beta > 2$
\Mevinv, as may be seen by the solid line in Fig.~\ref{lnzeta_sm148}.  We will therefore
use this method for the canonical quantities calculated below.

\subsubsection{Excitation energies} 
 
\begin{figure}[htb]  
\begin{center}  
\includegraphics[width=  9.1 cm]{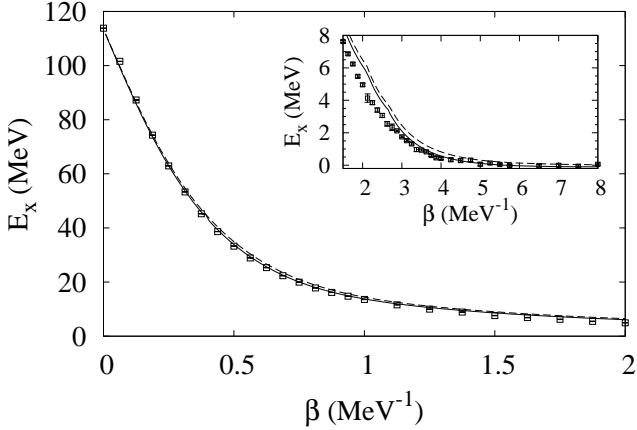} 
\caption{
Excitation energy of \sm~as a function of inverse temperature $\beta$  for
$0 < \beta < 2$ MeV$^{-1}$, comparing the grand canonical HFB energy in Eq.~(\ref{EHFB})
(dashed line) with
the approximate canonical energy in Eq.~(\ref{E0}) (solid line) 
and the SMMC results (open squares). The inset shows an expanded
energy scale.}
\label{evb_sm148}  
\end{center}  
\end{figure} 
 
The \sm~thermal excitation energy function $E_x$ vs.~$\beta$ in FTHFB is compared with the SMMC results in Fig.~\ref{evb_sm148}.  The HFB energy is shown by the dashed
line.   The excitation energies at $\beta=0$ are
nearly equal, differing only by the  missing correlation energy.
The two kinks in the HFB curve, visible in the inset, are associated
with the neutron and proton pairing phase transitions.  The HFB excitation energy
is higher than that of the SMMC, as to be expected from the 
higher limiting entropy of the grand canonical
ensemble.  We next calculate the approximate canonical projection of the HFB
energy using Eq.~(\ref{zeta-2}).  The result is closer to the SMMC for high $\beta$, 
We note that for $\beta < ~ 2$
\Mevinv (i.e., for temperatures above the pairing transitions) , the absolute HF and HFB energies coincide, so that the HF
excitation energy is lower than the HFB excitation energy by exactly the
amount of pairing correlation energy in the ground state.
 
\subsubsection{Entropies} 
 
\begin{figure}[htb]  
\begin{center}  
\includegraphics[width=  9.1 cm,angle=0,origin=c]{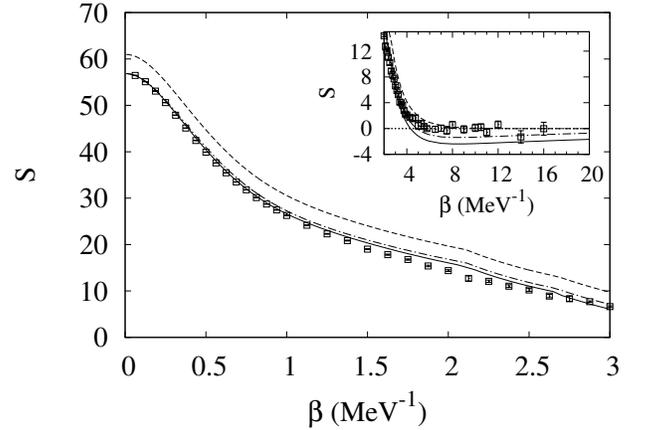} 
\caption{
Entropy functions of \sm:  the grand canonical HFB entropy (dashed line) and
the approximate canonical HFB entropy in Eq.~(\ref{S_c}) with $\zeta$
given by Eq.~(\ref{zeta-2}) (solid line) are compared with the SMMC entropy
(open squares).  The dashed-dotted line is the entropy associated with
the 3-D saddle-point point approximation, i.e., omitting the
$\beta \delta E$ term in Eq.~(\ref{S_c}).  
The
inset shows an expanded entropy scale at large $\beta$ values.}
\label{svb_hfb_sm148}  
\end{center}  
\end{figure}

The grand canonical HFB entropy (dashed line) and SMMC entropy (open
squares) functions in \sm~are shown in Fig.~\ref{svb_hfb_sm148} vs.~$\beta$. 
Their absolute values are set by integrating from the $\beta=0$ point, where
their respective values are given in Table I.  Both entropy functions
approaches zero at large $\beta$, confirming the sum rule (\ref{sr}).  
The neutron and proton pairing transitions are also visible as
kinks in the HFB entropy curve~\cite{Cv}.
 
We also show in Fig.~\ref{svb_hfb_sm148} the approximate canonical HFB
entropy in Eq.~(\ref{S_c}) where $\zeta$ is given by Eq.~(\ref{zeta-2}) in the discrete
Gaussian model. Since the particle-number fluctuations in FTHFB remain relatively large even at low temperatures, similar results for $\zeta$ are found in the saddle-point approximation Eq.~(\ref{zetasp}).

This approximate canonical entropy coincides with the SMMC at low values of
$\beta$ and overestimates the SMMC entropy around $\beta \sim 2$ \Mevinv~,
i.e., in the vicinity of the proton pairing transition.  At larger values of
$\beta$, for which a nonzero pairing condensate exists, the approximate canonical
entropy is in overall agreement with the SMMC entropy up to $\beta \sim 3.5$
\Mevinv~but at lower temperatures it becomes negative with a value of about
-2 at $\beta \sim 7$ \Mevinv~when the system reaches its HFB ground state.
We note that if $\zeta$
were to be calculated using the HFB particle-number fluctuations, the
large-$\beta$ entropy would have been even more negative at $\sim -4$ units.
A negative entropy at zero temperature is unphysical since there is only one state in the ensemble at
zero temperature and the entropy should be zero.  
The HFB ground state violates particle-number conservation;
the probability $\zeta^{-1}$ of having the proper proton and neutron numbers  
for \sm~at $\beta=7$ MeV$^{-1}$ is only $\sim 0.17$, hence
the unphysical negative entropy at low temperatures. 
 
To summarize the results of this section, we have not found a simple acceptable 
procedure to project the HFB onto a canonical ensemble if we require
the correct entropy at high temperature and an error of less that one
unit at low temperatures.  We will comment further
on this situation in the conclusion.

\subsubsection{Angular momentum fluctuations} 
 
Eq.~(\ref{DJ-HF}) which describes the angular momentum fluctuations in the FTHF approximation, has an additional contribution in FTHFB from a contraction in Wick's theorem that involves the anomalous density $\kappa$ 
\begin{eqnarray}\label{DJ-HFB} 
\langle (\Delta J_q)^2\rangle & = & \langle J_q^2 \rangle -\langle J_q\rangle^2 \nonumber \\ & = & \tr [j_q \, (1-\varrho) j_q \, \varrho] - \tr [j_q \,\kappa j^*_q \, \kappa^\ast]  \;. 
\end{eqnarray}
The additional  contribution is negative, leading to a reduction in the mean-square moment of the angular momentum.  This is just what one would
expect as an effect of the pairing correlations.
 
\begin{figure}[htb]  
\begin{center}  
\includegraphics[width= 9.1 cm]{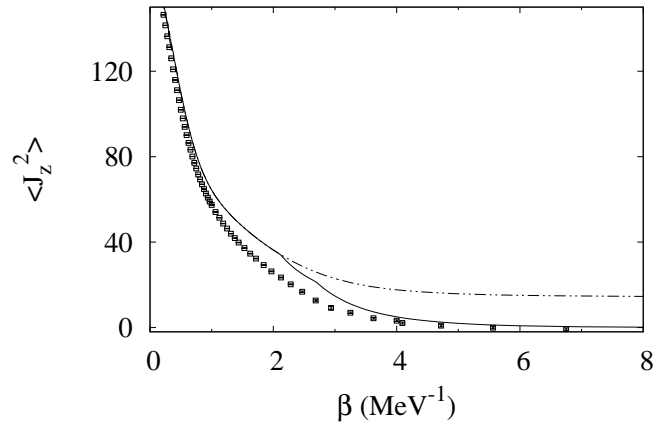} 
\caption{Mean square angular momentum $\langle J^2_z\rangle$ in \sm, comparing  the HFB results (solid line) with 
the SMMC results (open squares). The dashed-double dotted line corresponds to the spherical HF solution.  
} 
\label{jzvb_sm148}  
\end{center}  
\end{figure}

In Fig.~\ref{jzvb_sm148} we show the angular momentum fluctuations in \sm~as
a function of $\beta$.  The HFB solutions in \sm~is spherical at all
temperatures so all directions are equal and we only need examine one of
them.  We compare $ \langle J^2_z\rangle$ for HFB (solid line) with its SMMC
(open squares) values.  Below the pairing transition temperature, the HFB
values are strongly suppressed compared to the spherical HF solution (dashed
line), a known effect of pairing correlations. The SMMC values are further
suppressed compared with HFB, in particular in the vicinity of the pairing
transition. We observed substantial suppression also above the pairing
transition temperature, indicating the persistence of pairing correlations
in the SMMC results, even when the mean-field condensate no longer exists.

While the difficiency of the HFB around the phase transition is interesting
to see, the magnitude of the error is not large enough to be of concern in
calculating level densities.  

\subsubsection{State densities and level spacing} 
 
\begin{figure}[htb]  
\begin{center}  
\includegraphics[width= 9.1 cm,angle=0,origin=c]{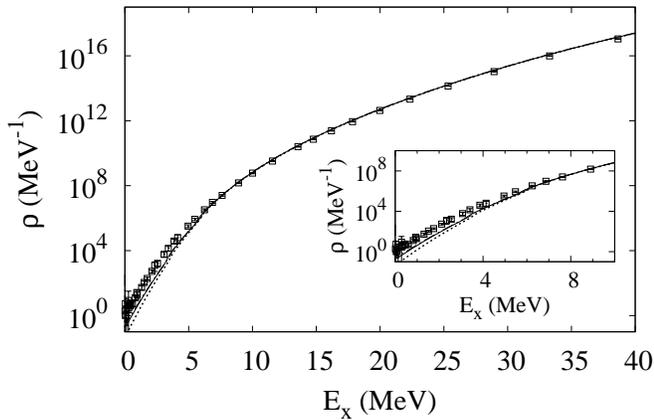} 
\caption{State densities in \sm.~The HFB density (solid line) calculated from Eqs.~(\ref{s2r}), 
(\ref{S_c}) and (\ref{zeta-2})  is compared with the SMMC state density (open squares). 
The dotted line is the HFB density with $\zeta$ calculated from the particle-number fluctuations, Eq.~(\ref{zeta}). The inset is an expanded scale at low excitation energies.} 
\label{rho_sm148}  
\end{center}  
\end{figure}

 In Fig.~\ref{rho_sm148} we show the HFB density of \sm~(solid
line) in comparison with the SMMC state density (open squares).  The HFB density
was calculated with the canonical saddle-point approximation (\ref{s2r}), taking
the canonical entropy from Eq.~(\ref{S_c}) and $\zeta$ from
Eq.~(\ref{zeta-2}).  The result is practically indistinguishable when the
$\delta E$ term in Eq.~(\ref{S_c}) is omitted. 
We observe good agreement between the HFB and SMMC densities for excitation 
energies above the pairing transitions $E_x \geq 7$ MeV. At those energies, 
the HFB solution coincides with the HF solution, and the only role of 
the pairing is to reset the origin of the excitation energy scale by the pairing 
correlation energy. This good agreement may be fortuitous in view of two 
compensating errors in the HFB: the missing correlation energy in the ground 
state resets the excitation scale to lower the level density, while the 
 many-body correlations increase the level density. 
This is seen as an increase in the effective mass of the 
quasi-particles~\cite{ga58}.  Beyond that, for attractive interactions, the  
RPA correlation energy further raises the level density~\cite{va96}. 
Nevertheless, we can take the present agreement as support for a popular model of the level density, namely the 
back-shifted Fermi gas. That model assumes that pairing correlations  only affect 
the origin of the excitation energy scale.   
 
At the same time, the calculated HFB level density is too small at low 
excitation energies.  As with the calculated canonical entropy coming out
negative, the problem relates to the violation of particle-number 
conservation in HFB.
 
Lastly we compute the neutron  
resonance spacing at threshold. For that we need the spin dependence of the level density.  We use 
Eqs.~(\ref{spectro}) and (\ref{PJ}), as was done for the SMMC, except that 
now $\rho(E)$ is taken to be the HFB density. This is justified since the 
HFB solution is spherical.  The spin cutoff factor $\sigma$ is taken from 
the HFB variance of $J_z$ (see Fig.~\ref{jzvb_sm148}).  As discussed 
previously, these fluctuations are larger in the HFB than in SMMC.  
However, this differences will affect the level density by less 
than a factor of two.  We find in HFB a neutron resonance spacing at the 
neutron threshold of 4.1 eV, in very good agreement with the SMMC value of 
3.7 eV (see Table III). 
 
\section{Conclusion and outlook}   
\label{conclusion} 
 
Our benchmarking of the finite-temperature HF and HFB approximations for level  
densities in heavy nuclei provides a quantitative assessment of the 
limitations of these mean-field theories, but also justifies their use 
under fairly broad sets of conditions.
We have emphasized the relation between the grand canonical and 
canonical statistics because the mean-field theories are  
formulated in the grand canonical ensemble, but the actual level densities 
and the SMMC theory used for benchmarking are canonical.  In the FTHF (which provides an appropriate mean-field approximation for a nucleus with weak pairing correlations), we found a simpler
way to approximate the projection from the grand canonical to the canonical ensemble when the particle-number fluctuation is small and the standard saddle-point approximation breaks down.
The corresponding formulas, Eqs.~(\ref{S_c}) and (\ref{zeta}), are accurate to much less than 
a unit of entropy, which is entirely acceptable in view of other 
sources of error.  However, we found no simple way to project the HFB to 
the canonical ensemble with accurate entropies at temperatures below the pairing transition.   
 We will come back to this later.   
 
A fundamental problem is how to treat broken symmetries. As has been 
long known, both the pairing and shape transitions are rather smooth in the finite-size nucleus, 
in contrast to the sharp phase changes that are present in the mean-field  
theory.  Otherwise, the 
issues that arise in the context of level densities are rather different for the shape and the pairing 
transitions.  The deformed shape of the  
nucleus \dy~is quite robust, persisting to excitation energy of $\sim 20$ 
MeV, so the change of shape with temperature can 
more or less be ignored in the statistical regime accessible with 
low-energy neutrons.  The HF state density of \dy~is too low by 
about an order of magnitude at low excitation energy, but the physics is 
easy to understand:  the HF is describing band heads and not the 
individual rotational states in each band.   
This may be seen in the calculated canonical entropy (see, e.g., inset of Fig.~\ref{svb-dy162}).  The SMMC entropy is $\sim 2-4$ units at 
$\beta \sim 5-10$ \Mevinv~due to the rotational band contribution. 
In contrast, the HF entropy is close to zero in that region. 
We attempted to take rotational band physics into account 
for level densities at small $J$ by treating each HF state as  
a rotational band head, and neglecting the rotational energies as 
in Ref.~\onlinecite{bj74}. 
This gave a level density at neutron threshold that is 
$\sim 5$ times larger than the SMMC benchmark. We conclude that a 
better treatment of rotational band structure
is required to achieve a good accuracy 
for applications.   
 
The HFB approximation for nuclei with strong pairing correlations is a  
well-established approximation for ground-state properties. Unfortunately, we found no 
simple way to project from the grand canonical to the canonical  
ensemble, due to the particle-number mixing that is inherent in  
the HFB wave function.  Furthermore, the odd-even energy  
staggering of paired systems requires that the variational principle must take 
into account the number parity of the ensemble.  According to 
Ref.~\cite{ba99}, the effects may persist in a nuclear context 
up to temperatures approaching the mean-field transition point.     
Fortunately for many applications, the 
pairing condensate disappears at rather low excitation energy, 
and the theory above this energy reduces to the HF approximation. The only effect of 
pairing correlations at these higher energies is to reset the excitation energy scale by adding 
the pairing correlation energy, as in the .  
well-known as the ``back-shift" 
models of level densities~\cite{ra97,eg05,ga13,po14}. 
  
In view of the comparative benchmarking of a deformed and a 
spherical nucleus, it is interesting to revisit the arguments 
presented by Bj\/ornholm, Bohr and Mottelson for effects of breaking 
the rotational symmetry on the level density~\cite{bj74}.  First of all, there is no 
visible effect when comparing the experimental level densities 
or the benchmark level densities of the two nuclei we examined, \dy~and \sm. 
In fact, we followed the prescription described in Eqs.~(8-10) of Ref.~\onlinecite{bj74} for extracting the  
level densities from band head densities, and obtained a level density that is too high.  Also, the benchmark entropies of both 
nuclei are very similar for $\beta$ in the range $\sim 1.5 - 3$ \Mevinv. 
We conclude that, except for the very lowest excitation energies, 
the deformation is much less important than commonly assumed. 
     
In a theory of the statistical properties of nuclei, an important source of 
error is calculating the baseline for the excitation energy of the thermal 
ensemble.  Any correlations in the ground state beyond mean-field theory  
will lower the base ground-state energy and thus tend to increase the  
excitation energy of a thermal ensemble. On the other hand, if similar 
correlations are also present at the excitation energies of interest,  
their effect will partially cancel the shift of the baseline.  
The more problematic 
correlations are those that are present in the ground state but 
are suppressed in the thermal ensemble at the excitation energies of 
interest.  A possible 
explanation of the too-high level density calculated for \dy~might be that 
the rotational energy is in this category.  On the other hand, 
we saw that the missing correlation energy of \sm~is similar to that of 
\dy~(see Table II).  It cannot change very much in \sm~and still keep good 
agreement with the SMMC. 
 
We conclude with some remarks on the possibilities of improving and 
extending the mean-field treatments discussed here, short of doing the full  
SMMC sampling. 
 
The mean-field theory is the first approximation in the systematic many-body 
perturbation theory.  The next approximation adds the second-order 
corrections to the energy or the grand potential $\Omega$.  In the 
infinite Fermi gas, these corrections increase the level density  
irrespective of the sign of the interaction; it would be interesting to  
determine if this is the case for the non-collective correlations within the 
shell model Hamiltonian spaces in use.     
 
The next systematic correction to mean-field theory is the random-phase approximation (RPA). 
It provides a powerful method to treat correlation energies, even in the 
presence of degeneracies that are associated with the broken symmetries.   
With some exceptions~\cite{va83,qu10}, the RPA has mostly been applied in  
the framework of the static path approximation (SPA)~\cite{ke81}. 
In early studies it was found that the ground-state energy obtained in the SPA  
was not accurate enough to be useful for setting the excitation energy  
scale for the level densities.  However, later 
model studies that included the RPA corrections to pairing have been quite  
successful~\cite{pu91,ro98} and the method has been applied to physical  
systems~\cite{ka06}.  We note also that the SPA+RPA with the
inclusion of number-parity projection describes well thermodynamic properties of superconducting 
nanoscale metallic grains~\cite{ne13}.  In the nuclear context, up to now there have not been 
any systematic study of 
the accuracy of the SPA+RPA in the framework of the 
shell model Hamiltonian such as the one used in the SMMC; we 
plan to examine it in the future.   
  
There is also an interesting recent study within the HF-BCS theory using the 
combinatorial method rather than the grand canonical ensemble 
~\cite{uh13}.  This method combines the IPM of the
the HF mean field together with the pairing energy from BCS 
with specific blocked orbitals.  This requires a large number of 
BCS calculations, but it was still possible the carry out a systematic survey of the level densities of deformed nuclei that 
extends up to the actinides.  We found that the IPM based on  
a deformed HF ground state was accurate to within one unit of  
entropy up to excitation energy of $\sim 10$ MeV in \dy. 
 
A final issue is the lack of 
a systematic theory simpler than the SMMC that applies equally well to 
deformed and paired spherical nuclei.
Ref.~\onlinecite{uh13} limits 
itself to deformed nuclei only; the global studies reported 
in Ref.~\onlinecite{hi06} and successor papers uses different 
formulas for spherical and deformed nuclei. In principle, 
the SPA could be a basis for 
more systematic theory.  However, that would require keeping explicit integrations 
over the two pairing fields (for protons and neutrons)  and the two intrinsic quadrupole fields.   
Furthermore, the problem of setting the excitation energy baseline remains an 
obstacle to using the SPA+RPA as a global theory. 
 
We hope that the present availability of realistic Hamiltonians and 
accurate SMMC calculations of their thermal and statistical properties will provide 
guidance to a renewed search for better methodologies for approximate theories that are less computationally intensive.  
 
\section{Acknowledgments}
This work was supported in part by the
U.S. DOE grant Nos. DE-FG02-91ER40608 and DE-FG02-00ER411132, and by Grant-in-Aid for Scientific Research (C) No. 25400245 by the JSPS, Japan.  The research presented here used resources of the National Energy Research Scientific Computing Center, which is supported by the Office of Science of the U.S. Department of Energy under Contract No.~DE-AC02-05CH11231. It also used resources provided by the facilities of the Yale University Faculty of Arts and Sciences High Performance Computing Center.

\section*{Appendix I: accuracy of the saddle-point approximation} 
 
Here we use a simple model to assess the accuracy of 
the saddle-point approximations for the state density in
Sec.~\ref{state-density} and 
the approximate
particle-number projections in Secs.~\ref{saddle-app} and \ref{discrete-gaussian} for the
entropy. The model describes independent fermions
populating equidistant energy levels.  The only parameter in the model (in the limit when the number of single-particle levels is large) is
the spacing of the single-particle states $\delta$.  The Hamiltonian of this
system is

\be\label{equidistant} 
H = \delta \sum_{i=0}^{\Omega-1} (i+1/2) 
a^\dagger_i a_i  
\ee 
where $\delta$ is the single-particle level spacing  and $\Omega$ is the total number of single-particle levels.  
 
We first show that the factorization (\ref{factorization}) of the grand canonical partition function is essentially exact in this model when  $T/ \delta$, the temperature in units of the single-particle mean-level spacing, is much smaller than both the number of particles $N$ and $\Omega-N$.  
The key observation is that under these conditions the canonical partition function $Z_c(\beta)$, calculated with respect to the ground-state energy of the $N$ particles (i.e., in terms of excitation energy) is {\em independent} of $N$.  Changing $N$ shifts the Fermi energy but since the single-particle spectrum is invariant under such a shift, the particle-hole excitations remain unchanged. A typical particle-hole excitation energy is of order $T$  so the number of excited particles is typically smaller than $N$ (under the above conditions).  
 
The canonical partition function of a system with ground-state energy $E_0$ is given by $e^{-\beta E_0} Z_c(\beta)$ where $Z_c(\beta)$ is the partition calculated using the excitation energies.  The $N$ particle ground-state energy of the Hamiltonian (\ref{equidistant}) is given by $E_0= N^2 \delta /2$, and thus the $N$ particle canonical partition is 
\be\label{Z-c-N} 
Z_c(\beta,N) = e^{-\frac{1}{2} \beta N^2 \delta} Z_c(\beta) \;. 
\ee 
We expand the grand canonical partition function $Z_{gc}(\beta,\alpha)$ for a value $\alpha=\alpha_0$ that gives an average number of particles $N_0$ 
\be\label{a-0} 
\alpha_0 =  \beta N_0  \delta \;, 
\ee 
 and use Eq.~(\ref{Z-c-N})  to find 
 \be 
Z_{gc}(\beta,\alpha_0) \approx \sum_N  e^{\alpha_0 N -\frac{1}{2} \beta N^2 \delta }  Z_c(\beta) \;. 
\ee 
The quasi-equality ``$\approx$" is a reminder that the formula is valid in  
for $T/\delta \ll N, \Omega-N$. 
Using (\ref{a-0})  we have 
\be 
Z_{gc}(\beta,\alpha_0) \approx  \sum_N  e^{-\beta \delta {(N-N_0)^2 \over 2}} e^{\frac{1}{2} \beta {N_0^2} \delta } Z_c(\beta) \;. 
\ee 
With the help of Eq.~(\ref{Z-c-N}) for $N=N_0$,  we can rewrite the last relation in the form 
\be 
Z_{gc}(\beta,\alpha_0) e^{-\alpha_0 N_0}  = Z_c(\beta,N_0)   \left(\sum_N  e^{-\beta \delta {(N-N_0)^2 \over 2}}\right)  
\ee 

This relation describes the factorization (\ref{factorization}) of the grand
canonical partition. The quantity in parenthesis is the partition function
$\zeta$ of the discrete Gaussian model [see Eq.~(\ref{zeta})], provided that
the particle-number variance is $(\beta\delta)^{-1}$.  Indeed

\begin{eqnarray}
\langle (\Delta N)^2\rangle &=& \sum_{m=0}^{\Omega}  f_m (1-f_m) \nonumber \\ & \approx & \int_0^{\Omega} dm {e^{\beta m \delta -\alpha} \over(e^{\beta m \delta -\alpha} +1)^2} = {1 \over \beta\delta} \;,
\end{eqnarray}
where we have used $\beta N_0 \delta \gg 1$ and $\beta (\Omega - N_0)\delta \gg 1$.  
 
It is instructive to see how the particle-number projection works with a numerical example. 
We take the ensemble is a finite space, $(\Omega, N) = (40,20)$, chosen to have dimensions 
comparable to those we dealt with in the text. 
 
\begin{figure}[htb] 
\begin{center}  
\includegraphics[width= 9.1 cm]{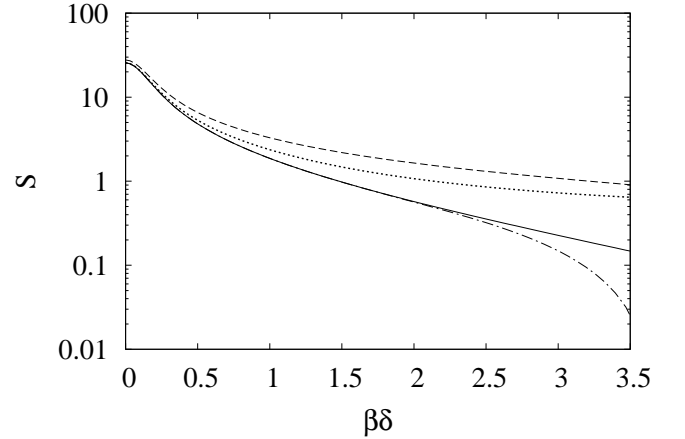}  
\caption{Entropy of the $(\Omega,N)=(40,20)$ model as a function of $\beta$. 
See text for explanation. 
} 
\label{Svb_ipm_40}  
\end{center}  
\end{figure}

The exact canonical entropy $S_c(\beta)$ (obtained by using exact particle-number projection) is shown 
in Fig.~\ref{Svb_ipm_40} by the solid line.  It starts 
at $S(0) = \ln { 40 \choose 20} = 25.65$ and approaches zero at large $\beta$. 
 
We next turn to the grand canonical ensemble, in which we fix the chemical
potential at each value of $\beta$ to get the desired particle number in the
ensemble average. The grand canonical entropy for the $(40,20)$ model is shown as the dashed
 curve in Fig.~\ref{Svb_ipm_40}. Here the starting entropy from
Eq.~(\ref{Sgci0}) is 27.73, larger than the canonical entropy by 2.08 units. 
The approximate reduction to the canonical ensemble is carried out by
Eq.~(\ref{S_c}). The result is shown as the dashed curve in
Fig.~\ref{Svb_ipm_40}.  It is accurate to within 0.1 of a unit over entire
range of $\beta$. We also show for comparison the entropy calculated 
with Eq.~(\ref{S_c})
without the $\delta E$ correction (dotted line).  It is much
less accurate, differing from the exact value by more than a half-unit
for $\beta\delta > 1$.
 
\begin{figure}[htb] 
\begin{center}  
\includegraphics[width= 9.1 cm]{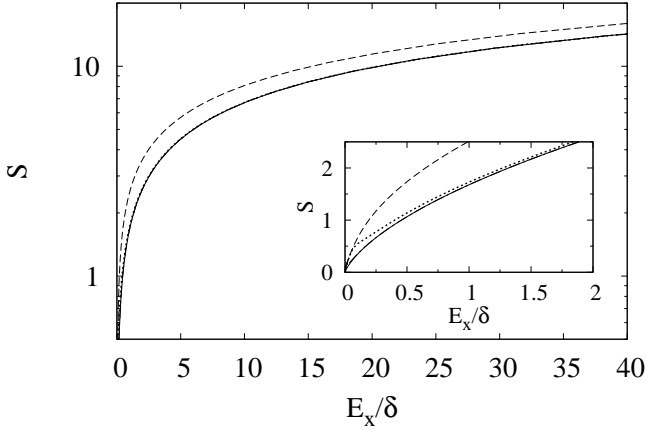}  
\caption{Entropy of the $(\Omega,N)=(40,20)$ model as a function of excitation energy $E_x/\delta$. 
See text for explanation. 
} 
\label{Sve_ipm_40}  
\end{center}  
\end{figure}

The entropies of Fig.~\ref{Svb_ipm_40} are reploted in Fig.~\ref{Sve_ipm_40}
as a function of excitation energy (in units of $\delta$) $E_x/\delta$.  The entropy shown by the 
dashed line is the approximate canonical entropy of Eq.~(\ref{S_c}) that includes the contribution from $\delta E$; in this contribution is omitted in the entropy shown as the dotted line. The two are very close 
except at low excitation energies (see inset). 
The difference of the approximate canonical entropy from the true canonical
entropy (solid line) can hardly be seen in the figure.
 
\begin{figure}[h!]  
\begin{center}  
\includegraphics[width= 9.1 cm]{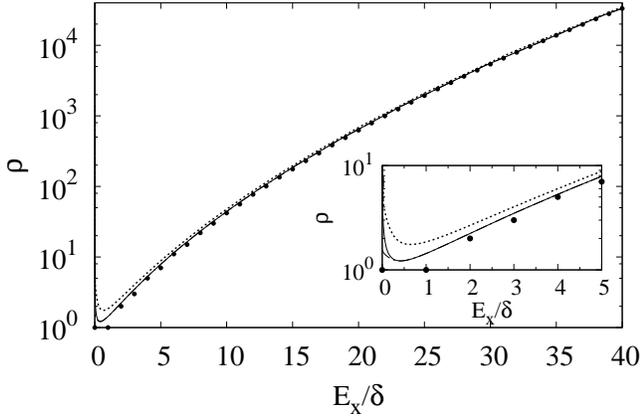} 
\caption{State density for the Hamiltonian (\ref{equidistant}) with $(\Omega,N) = 
(40,20)$. See text for explanation. 
} 
\label{rho_hist}  
\end{center}  
\end{figure} 
 
We next turn to the state density itself.  The excitation energy spectrum is $n\delta$ ($n=0,1,2,\ldots$), and the state density  for $N$ particles  is given by 
\be 
\label{rN} 
\rho_N(E_x) = \sum_{n=0} \delta(E_x -n\delta) \;a(n) 
\ee 
for $n <  \min(N,\Omega-N)$.  Here $a(n) = 1,1,2,3,5,7,11,...$ is  the well-known partition of the integer $n$~\cite{euler,AB}. 
The saddle-point approximation for the state density is also 
well-known~\cite{AB} and it has been shown to be  accurate enough for our 
purposes for $n~>2$~\cite{er60}.   
 
Fig.~\ref{rho_hist} shows as solid circles the state density as the number of states within energy bins of width 
$\delta$ (i.e., the numbers $a(n)$). The ground state and first excited state are unique, and 
then there is an increasing density up to half the maximal energy.  For comparison we also show by solid line the level density
calculated from (\ref{s2r}) using the canonical energy and entropy obtained
by exact particle-number projection.

The state density in the standard saddle-point approximation is shown by the
dotted line, while the state density of Eq.~(\ref{s2r}), in which the
saddle-point prefactor is calculated from $d E_\zeta /d\beta$ (see Eq.~(\ref{E0})) gives the dashed-dotted line.  This curve is hardly distinguishable from the solid line and improves the agreement with the
exact result at low excitation energies. We observe that the improved
saddle-point expression (\ref{S_c}) and (\ref{E0}) is accurate to better than $10\%$ to
energies as low as $2\delta$.  Since the low-lying region of the spectrum would
be calculated by explicit methods anyway, we conclude that the improved
saddle-point approximation is entirely adequate for statistical purposes.
 
\section*{Appendix II: thermodynamical consistency of the HF and HFB approximations} 
 
The key consistency condition of a finite-temperature theory in the grand
canonical ensemble is the relation between $S_{gc}$ and $E_{gc}$ given
by the equation analogous to Eq.~(\ref{S}) for the canonical ensemble.
It can be easily derived from the relations (\ref{3D-saddle}) satisfied by the first logarithmic derivatives
of $Z_{gc} = \Tr \exp(-\beta \hat H + \alpha \hat N)$. 
 However, these derivatives are more subtle in
the case of a mean-field Hamiltonian because the effective Hamiltonian in the density operator depends on the temperature and chemical potentials.
 
Here we prove that similar relations Eqs.~(\ref{E-N-HF1}) are indeed valid in the FTHF  
approximation. The proof follows from the fact that the theory is  
derived from the variational principle 
for the grand potential $\Omega$~\cite{ta81,go81}. 
We write expression for the HF grand  potential $\Omega_{HF}$ in the form 
\be\label{Omega_HF} 
-\ln Z_{HF} = \beta\Omega_{HF} =  \beta E_{HF} - S_{HF} - \alpha N_{HF}\;. 
\ee 
where $E_{HF},S_{HF}$ were defined in Eqs.~(\ref{E-HF}) and (\ref{S-HF}). 
 
The derivative of Eq.~(\ref{Omega_HF}) with respect to $\beta$ has a 
contribution $E_{HF}$ from the explicit dependence on $\beta$.  However, 
there is in principle also a contribution from the implicit dependence on 
$\beta$ in $E_{HF},S_{HF}$ and $N_{HF}$.  To see that they vanish we go back  
to the many-body uncorrelated density matrix $\hat D_{HF}$ that was the trial density of the 
variational principle.  Taking that as the fundamental variable, the 
relevant derivative is 
\begin{eqnarray}\label{dlnZ} 
-{\partial \ln Z_{HF}\over \partial \beta} \Big |_\alpha & = &  
{\partial (\beta\Omega_{HF})\over \partial \beta}\Big |_\alpha \nonumber \\
&= & E_{HF} + { \delta (\beta \Omega_{HF})\over \delta \hat D_{HF}}\Big |_{\beta,\alpha} {\partial\hat  D_{HF} \over \partial \beta} \;. 
\end{eqnarray} 
Since $\hat D_{HF}$ is a variational solution at fixed $\beta$ and $\alpha$, it follows that  
\be 
 {\delta (\beta \Omega_{HF})\over \delta \hat D_{HF}} \bigg |_{\beta,\alpha}=0 \;,  
 \ee 
and the second contribution on the right-hand side of Eq.~(\ref{dlnZ}) 
vanishes.  The thermodynamic consistency of the finite-temperature HFB can 
be proven in the same way.


\begin{thebibliography}{99} 
\bibitem{go81} A.L. Goodman, Nucl. Phys. {\bf A352} 30 (1981).  
\bibitem{ta81} K. Tanabe, K. Sugawara-Tanabe and H.J. Mang, Nucl. Phys. 
{\bf A357} 20 (1981). 
\bibitem{ke81} A.K.~Kerman and S.~Levit, Phys. Rev. C {\bf 24}, 1029 (1981); 
A.K.~Kermin, S~Levit and T.~Troudet, Ann. Phys. {\bf 148} 436 (1983). 
\bibitem{va83} D. Vautherin and N. Vinh Mau, Phys. Lett. B {\bf 120}, 261 
(1983); 
N. Vinh Mau and D. Vautherin, Nucl. Phys. A {\bf 445}, 245 (1985). 
\bibitem{la93} G.H.~Lang,  C.W.~Johnson, S.E.~Koonin, and W.E.~Ormand, Phys. Rev. C {\bf 48}, 1518 (1993).
\bibitem{al94} Y.~Alhassid, D.J.~Dean, S.E.~Koonin, G.H.~Lang, and W.E.~Ormand, 
Phys. Rev. Lett. {\bf 72}, 613 (1994). 
\bibitem{na97} H. Nakada and  Y. Alhassid, Phys. Rev. Lett.~{\bf 79}, 2939 (1997).
\bibitem{al99} Y. Alhassid, S. Liu and H. Nakada, Phys. Rev. Lett.~{\bf 83}, 4265 (1999).
\bibitem{BM} A. Bohr and B. Mottelson, {\it Nuclear Structure},  
 Vol. I, (Benjamin, 1975). 
\bibitem{det} The equivalence follows directly from the identity
for symmetric matrices $A$: 
$ \det A =  \det B ( a_{11} - \vec a_1 \cdot B^{-1} \cdot \vec a_1)$,
where $a_{11}$ is the element of $A$ in the first row and column, 
$B$ is the minor obtained from $A$ by striking out the first row and column,
and  $\vec a_1|_{i} = a_{1,i+1}$.
\bibitem{er60} T. Ericson, Advances in Physics {\bf 6}, 425 (1960). 
\bibitem{sign} Small bad-sign interaction terms can be treated using the extrapolation method of Ref.~\cite{al94}.  
\bibitem{or94} W.E.~Ormand, D.J.~Dean, C.W.~Johnson, et al., Phys. Rev. C 
{\bf 49}, 1422 (1994). 
\bibitem{al08}  Y. Alhassid, L. Fang and H. Nakada, Phys. Rev. Lett. {\bf 101}, 082501 (2008). 
\bibitem{oz13} C.~\"{O}zen, Y.~Alhassid, and H. Nakada, Phys. Rev. Lett. {\bf 110}, 042502 (2013). 
\bibitem{RIPL}  The RIPL3 database containing tables of experimental 
    level densities can be accessed at 
    {\tt https://www-nds.iaea.org/RIPL-3/resonances/}. 
\bibitem{es93} C. Esebbag and J.L.~Egido, Nucl. Phys. A {\bf  552}, 205 (1993). 
\bibitem{ro98}  R. Rossignoli, N. Canosa, and P. Ring, Phys. Rev. Lett. 
{\bf 80}, 1853 (1998). 
\bibitem{ka06} K. Kaneko, M. Hasegawa, et al., Phys. Rev. C {\bf 74}, 024325 
  (2006); K. Kaneko and A. Schiller, Phys. Rev. C {\bf 76}, 064306 (2007). 
\bibitem{ga13} D. Gambacurta, D. Lacroix, and N. Sandulescu, Phys. Rev. 
C {\bf 88}, 034324 (2013). 
\bibitem{bj74} S. Bj\o rnholm, A. Bohr and B.R.~Mottelson,  
{\it Physics and Chemistry of Fission 1973} (IAEA, Austria, 1974), p. 367. 
\bibitem{be14} G.F.~Bertsch and L.M.~Robledo, Comp. Phys. Comm. {\bf 185},
3406 (2014). 
\bibitem{Cv}  The mean-field phase transition can be seen even 
more clearly as a discontinuity in the heat capacity 
$C = \beta d S / d\beta$. 
\bibitem{ga58} V. M. Galitskii, Sov. Phys. JETP {\bf 34}, 151 (1958). 
\bibitem{va96} D. Vautherin, Adv. Nucl. Phys. {\bf 22}, 123 (1996). 
\bibitem{ba99} R. Balian, H. Flocard, and M. Veneroni, Phys. Rep.  
{\bf 317}, 251 (1999). 
\bibitem{ra97}  T. Rauscher, F.-K. Thielemann, and K.-L.~ Kratz, 
Phys. Rev. C {\bf 56}, 1613 (1997). 
\bibitem{eg05} T. von Egidy and D. Bucurescu, Phys. Rev. C {\bf 72}, 044311 
(2005). 
\bibitem{po14} I. Poitoratska, et al., Phys. Rev. C {\bf 89}, 054322 (2014). 
\bibitem{qu10} N. Quang Hung and N. Dinh Dang, Phys. Rev. C {\bf 81}, 
057302 (2010). 
\bibitem{pu91} G.~Puddu, P.F.~Bortignon, and R.A.~Broglia, Ann. Phys. (NY) 
{\bf 206}, 409 (1991). 
\bibitem{ne13} K.N. Nesterov and Y. Alhassid, Phys. Rev. B {\bf 87}, 014515 (2013), and references therein.  
\bibitem{uh13} H. Uhrenholt, S. \AA berg, A. Bobrowolski, et al., Nucl. 
Phys. A {\bf 913}, 127 (2013). 
\bibitem{hi06}  S.~Hilaire and S.~Goriely, Nucl. Phys. A {\bf 779}, 63 
(2006). 
\bibitem{euler} L. Euler (1753), cited by Ericson~\cite{er60}. 
\bibitem{AB} M. Abramowitz and I. Stegun, {\it Handbook of Mathematical 
Functions}  (National Bureau of Standards, 1964), Sec.~24.2.1. 

\end{thebibliography}
\end{document}